\documentclass[aps,pra,reprint,groupedaddress]{revtex4-2}
\usepackage{graphicx}% Include figure files
\usepackage{dcolumn}% Align table columns on decimal point
\usepackage{bm}% bold math
\usepackage{hyperref}% add hypertext capabilities
\usepackage{amsmath}
\usepackage{amssymb}
\usepackage{mathtools}
\usepackage{mathrsfs}
\usepackage{xcolor}
\usepackage{dsfont}
\usepackage{physics}
\usepackage{lipsum}
\usepackage{comment}
\usepackage[normalem]{ulem}
\usepackage{booktabs}

\begin{document}

\title{Master equation for systems interacting with linearized gravity }


\author{Oliviero Angeli}
\email{oliviero.angeli@phd.units.it}
\affiliation{Department of Physics, University of Trieste, Strada Costiera 11, 34151 Trieste, Italy}
\affiliation{Istituto Nazionale di Fisica Nucleare, Trieste Section, Via Valerio 2, 34127 Trieste, Italy}

\author{Anirudh Gundhi}
\email{anirudh.gundhi@units.it}
\affiliation{Department of Physics, University of Trieste, Strada Costiera 11, 34151 Trieste, Italy}
\affiliation{Istituto Nazionale di Fisica Nucleare, Trieste Section, Via Valerio 2, 34127 Trieste, Italy}

\author{Angelo Bassi}
\email{abassi@units.it}
\affiliation{Department of Physics, University of Trieste, Strada Costiera 11, 34151 Trieste, Italy}
\affiliation{Istituto Nazionale di Fisica Nucleare, Trieste Section, Via Valerio 2, 34127 Trieste, Italy}

\date{\today}

\begin{abstract}
We investigate the open quantum  dynamics of a system of two masses interacting with an environment of linearized gravitational waves. We formulate the analysis  in terms of the observable proper distance between the two masses, and show that the canonical variables obtained from the standard Lagrangian, expressed in terms of the Fermi normal coordinates, are not suitable for an effective description of the system. We resolve this issue through a unitary transformation that provides a physically meaningful system--environment decomposition and derive the master equation to leading order in $G$. Its dissipative sector reproduces the classical energy loss due to gravitational-wave emission, while the noisy contributions suppress coherences between states with different mass quadrupole, or effectively, different proper separations. In the regime where the proper distance can be described by considering small quantum fluctuations around  an average distance $l_0$, the dynamics reduces to a Caldeira--Leggett-type equation, with a decoherence rate dependent on the baseline length $l_0$. 
\end{abstract}

\maketitle
\section{Introduction}
Gravitational decoherence refers to the decay of quantum superpositions of an object, due to its interaction with the gravitational degrees of freedom, such as those ascribed to gravitational waves. This phenomenon has received significant attention over the years~\cite{anastopoulos1996quantum,breuer2009metric,anastopoulos2013master,blencowe2013effective,oniga2016,oniga2017quantum,asprea2021gravitational,asprea2021letter, malcolm2025signatures, parikh2021signatures,parikh2020noise,cho2025non, torovs2024loss,sharifian2024open} (See also the reviews~\cite{anastopoulos2022gravitational,bassi2017gravitational}) as a potential tool to gain further insights into the interplay between gravity and quantum theory. Indeed,  decoherence  depends on the form of the interaction Hamiltonian~\cite{paz1993environment}; therefore, from the details of the effective dynamics of the system, one can learn about how gravity couples to quantum matter. 

In light of potential experimental verifications, the weak field limit where gravity is linearized suffices for all practical purposes. Even though the classical action corresponding to linearized gravity is well-known, the suitable Hamiltonian for open quantum system dynamics becomes non-trivial to derive. The challenges are connected to the central tenet of decoherence theory: the system and environmental degrees of freedom have to be unambiguously separated, so that the partial trace over the environmental degrees of freedom can be performed, after which what remains is the effective dynamics for the system of interest, defined in terms of the system variables only.  In the present context, this poses two problems.

First, the gravitational degrees of freedom, namely the metric perturbation, contribute to defining the system observables such as the physical distance between two masses. If they  are traced out, they are not available anymore for computing such  physical quantities. This issue has been addressed in the literature by expressing the system degrees of freedom from the start in terms of the so-called Fermi normal coordinates \cite{malcolm2025signatures, parikh2021signatures,parikh2020noise,cho2025non, torovs2024loss,sharifian2024open}, which locally refer to physical distances, rather than the Cartesian coordinates \cite{blencowe2013effective,anastopoulos2013master, breuer2009metric,oniga2016,oniga2017quantum,asprea2021gravitational,asprea2021letter,anastopoulos1996quantum}, where the latter require the metric for computing physical quantities.

However, even after expressing the action in terms of the Fermi coordinates, a second non-trivial issue for having  well-separated and physically meaningful S-E variables arises in relation to the conjugate momentum of the system obtained via the standard Legendre transformation. This is needed in order to derive the Hamiltonian from the action, which in turn is required for canonical quantization. In a nutshell, the issue is that the conjugate momentum, while mathematically referring to the system alone, being by definition the conjugate of the system's position, operationally it depends also on the environment. To see how this is the case, let us briefly consider the paradigmatic example of electrodynamics. 

Consider the simple case of a non-relativistic particle interacting with the electromagnetic field. In the Coulomb gauge, the Lagrangian is given by~\cite{TannoudjiPartOneChapterTwo} $L_{\text{\tiny QED}} = \frac{1}{2}m\dot{\mathbf{x}}_e^2+\int\!d\mathbf{x} \mathcal{L}_{\text{\tiny{EM}}}-e\dot{\mathbf{x}}_{e}\mathbf{A}_{\perp}$.
Here, $\mathbf{x}_e$ denotes the position of the charged particle,  $\mathcal{L}_{\text{\tiny{EM}}}$ is the Lagrangian density for the free EM field, and $\mathbf{A}_{\perp}$ is the transverse vector potential. The conjugate momentum to $\mathbf{x}_e$ is $\mathbf{p} =m\dot{\mathbf{x}}_e -e\mathbf{A}_{\perp}$. If the electromagnetic field is treated as the environment and traced away, one is left with a master equation for the particle, which is function of  $\mathbf{x}_e$ and $\mathbf p$, typically a sum of double (anti-)commutators of these operators. This is consistent from the mathematical point of view, but problematic from the physical point of view, since the theoretical predictions about (functions of) $\bf p$ are not connected to experimentally accessible quantities: the position and velocity of the particle. This is because, as per the previous relation $\mathbf p = m\dot{\mathbf{x}}_e -e\mathbf{A}_{\perp}$, $\mathbf p$ is not a system observable alone, since it also involves $\mathbf{A}_{\perp}$, which is supposed not to be accessible as it has been traced away. See Ref.~\cite{EqLagrangians2026} for a full analysis of this problem.

In the present case of the gravitational interaction, the situation is even more complicated, and is analyzed in detail in Secs.~\ref{sec:FNC} and~\ref{sec:oqs}. The bottom line is that the conjugate variables and, as a consequence, the form of the Hamiltonian, have to be chosen appropriately in order for the trace to yield a mathematically consistent and physically meaningful master equation for the system. This is one of the main contributions of this work. 

The other  contribution concerns the master equation which is derived in Sec.~\ref{sec:pme}, after expressing the Hamiltonian in terms of the appropriate conjugate variables in Sec.~\ref{sec:Hamiltonian}. The master equation so obtained, using the influence functional formalism~\cite{feynman2000theory,calzetta2009nonequilibrium}, matches the classical expression for energy loss due to emission of gravitational bremsstrahlung, and predicts decoherence of superpositions corresponding to different proper distances between the two masses. Due to the form of the Lagrangian, expressed in terms of the Fermi coordinates, the master equation involves double commutators of the proper distances squared and, thus, has a different functional form compared to the well-known Caldeira-Leggett equation~\cite{caldeira1983path} which instead involves a double commutator of the (linear) position operator. In Sec.~\ref{sec:pme}, we consider the case where the relevant system degrees of freedom can be treated as small quantum fluctuations around a classical value, and show that in this limit the master closely resembles the Caldeira-Leggett equation.

The article is organized as follows. In Sec.~\ref{sec:FNC}, following the recent literature related to gravitational decoherence~\cite{parikh2021signatures,parikh2020noise,parikh2025quantum,kanno2021noise,cho2022quantum,moreira2025graviton,torovs2024loss,sharifian2024open,cho2025non,cho2023graviton}, we consider a system of two masses and adopt the Fermi-normal coordinates which describe the system in terms of proper distances. In Sec.~\ref{sec:oqs},  we show that the Lagrangian obtained within this description is inadequate for an open quantum system treatment~\cite{EqLagrangians2026}, since relevant system variables, like the velocity, cannot be captured at the level of the reduced density matrix. We overcome this issue in Sec.~\ref{subsec:UT} by identifying the most adequate S-E split, as defined by the unitary transformation that we put forward. In Sec.~\ref{sec:pme}, 
 we derive, via a perturbative calculation to $O(G)$, the Markovian master equation for the reduced evolution of the system. In Sec.~\ref{Sec:lin}, we consider the case where the quantum fluctuations of system are small, and can thus be described within the linearized limit. Finally, in Sec.~\ref{sec:deco}, we analyze the predictions of the master equations, the general and the linearized one, in regards to decoherence and show that coherence is lost between system-states superposed over different proper distances, as one would expect.

\section{The Lagrangian}\label{sec:FNC}
We consider weak gravitational fields, for which the spacetime metric $g_{\mu\nu}$ can be written as
\begin{equation}\label{smallg}
    g_{\mu\nu}(x) = \eta_{\mu\nu} + h_{\mu\nu}(x)\,.
\end{equation}
Here, $\eta_{\mu\nu} = \text{diag}(-1,1,1,1)$ is the Minkowski metric and $h_{\mu\nu}(x)$ are weak perturbations propagating on top of it.  General relativity is invariant under general coordinate transformations, from which   $h_{\mu\nu}$ inherits several properties and constraints. In particular, under an infinitesimal coordinate transformation $x^\mu \rightarrow x^\mu + \chi^\mu(x)$, the perturbation $h_{\mu\nu}$ transforms as
\begin{equation}\label{pre:gauge}
h_{\mu\nu} \rightarrow h_{\mu\nu}
- \partial_\mu \chi_\nu
- \partial_\nu \chi_\mu.
\end{equation}
Different choices of the functions $\chi_\mu$ correspond to different gauges, associated with the freedom to choose coordinates, albeit compatibly with the perturbative decomposition of Eq.~\eqref{smallg}, which implies $|\partial_\nu \chi_\mu|\ll 1$. 
The gauge freedom of Eq.~\eqref{pre:gauge} highlights that the coordinates by themselves do not carry a physical meaning.

When gravitational effects are relevant, coordinate distances need not coincide with physical, proper distances that can be measured. Although one may work in any coordinate chart, proper distances and proper time intervals are reconstructed through the metric by integrating the line element
\begin{equation}
ds^2=(\eta_{\mu\nu}+h_{\mu\nu})dx^\mu dx^\nu 
\end{equation}
along a given trajectory. In the present analysis, however, the gravitational perturbation $h_{\mu\nu}$ is treated as an environmental degree of freedom and will be integrated out. Therefore, if the reduced system was parametrized only by coordinate separations, the corresponding proper distances could not be reconstructed from the reduced state alone after tracing out the gravitational field. Then, for the reduced evolution variables to be meaningful, it is paramount that the system is parametrized from the outset in terms of proper, physical distances~\cite{malcolm2025signatures}. 

Following the recent literature on graviton noise~\cite{parikh2020noise,parikh2025quantum,cho2022quantum,cho2023graviton,kanno2021noise}, this can be achieved locally by writing the action in terms of the so-called Fermi Normal Coordinates (FNCs)\cite{manasse1963fermi,marzlin1994fermi} for a system of two massive objects. In FNCs, the origin is set along the trajectory of one of the two masses,  while the other object is labeled with the geodesic deviation vector $\xi^i$, which then encodes the proper distance between the two masses. The metric in the FNCs so defined, expanded to second order in the geodesic deviation vector, is given by \cite{marzlin1994fermi,rakhmanov2014fermi}
\begin{align}\label{III_TTmetric}
    g_{00} &= -1 - R_{0i0j}(0,t)\xi^i\xi^j\,,\nonumber\\
    g_{0i} & = -\frac{2}{3}R_{0jik}(0,t)\xi^j\xi^k\,,\\
    g_{ij} & = \delta_{ij} - \frac{1}{3}R_{ikjl}(0,t)\xi^l\xi^k\,.\nonumber
\end{align}
By construction, the Riemann tensor $R_{\mu\nu\lambda\sigma}$ in Eq.~\eqref{III_TTmetric} is evaluated along the reference geodesic, i.e., at the origin of the Fermi frame. The metric in Eq.~\eqref{III_TTmetric} is  the expansion of the Fermi-frame metric up to second order in the geodesic deviation vector $\xi^i$. Higher-order terms contain derivatives of the Riemann tensor, evaluated on the same geodesic. 
To estimate the size of these terms, consider a plane-wave  $h_{ij}(t,\mathbf{x}) \sim e^{i(\mathbf{k}\cdot\mathbf{x}-\Omega t)}
$ with $\Omega=c|\mathbf{k}|$ and $\lambda=\frac{2\pi}{|\mathbf{k}|}.$
Since the Riemann tensor is linear in second derivatives of $h_{\mu\nu}$, one has, schematically, 
\begin{equation}
R_{\mu\nu\lambda\sigma}\sim \partial^2 h\sim k^2 h\sim \frac{h}{\lambda^2}\,.
\end{equation}
Meanwhile the derivatives of the Riemann tensor scale as
$\partial R_{\mu\nu\lambda\sigma}\sim k^3 h\sim h/\lambda^3,$ and $
\partial^2 R_{\mu\nu\lambda\sigma}\sim k^4 h\sim h/\lambda^4$. 
Therefore, the retained quadratic terms in Eq.~\eqref{III_TTmetric} scale as
$\left(\xi/\lambda\right)^2$, whereas the first omitted terms scale as $\left(\xi/\lambda\right)^3$. Thus, to consistently use Eq.~\eqref{III_TTmetric}, we will assume that all the wavelength modes of the metric perturbation satisfy $\xi^i\ll \lambda$ (akin to the well-known dipole approximation in electrodynamics). Truncating the series expansion to second order in $\xi^i$ thus provides a natural UV cutoff $\Omega_\text{max}$ upto which the calculations can be trusted, such that 
\begin{align}\label{eq:cutoff}
\Omega_\text{max}\ll 2\pi c/\xi \simeq 10^9(m/\xi)s^{-1}\,,
\end{align}
where $\Omega_{\text{max}} := k_{\text{max}} c$, and $k_{\text{max}}$ is the wavenumber corresponding to the shortest admissible wavelength mode of the metric perturbation. 

Using Eq.~\eqref{III_TTmetric}, the action for the two masses, in FNCs, is given by~\cite{kanno2021noise}
\begin{align}
    S_\text{mat} =& -mc^2\int dt \sqrt{-\frac{1}{c^2}g_{\mu\nu}(t,\xi)\dot \xi^\mu\dot \xi^\nu} \nonumber \\
    &\simeq \int dt \Big(\frac{m}{2}\delta_{ij}\dot\xi^i\dot\xi^j-\frac{mc^2}{2}R_{i0j0}\xi^i\xi^j\Big)\,,
\end{align}
where $\xi^{\mu} = (ct,\xi^{i}(t))$, $m$ is the reduced mass of the system, and where we have discarded an irrelevant constant. 

Within the linearized limit of the gravitational field, which is what we restrict ourselves to, the Riemann tensor is gauge-invariant~\cite{maggiore2008gravitational} and hence can be evaluated in any gauge. For convenience, we  choose the well-known Transverse-Traceless (TT) gauge where 
\begin{equation}\label{II_TTGauge}
    h_{\mu 0} = 0\,,\qquad \partial^i h_{ij} = 0\,,\qquad h^i_i = 0 \,,
\end{equation}
such that~\cite{maggiore2008gravitational,carroll2019spacetime} 
\begin{equation}\label{III_Riemann}
    R_{i0j0} =-\frac{1}{2c^2}\ddot h_{ij}(0,t)\,.
\end{equation} 
Thus, the full matter-gravity action reads
\begin{align}\label{III_MatActionPol}
    S =\int dt \Big(\frac{m}{2}\delta_{ij}\dot\xi^i\dot\xi^j+\frac{m}{4}\ddot h_{ij}(0,t)\xi^i\xi^j\Big) +S_\text{free}\,,
\end{align}
where $S_\text{free}$ is the free graviton action
\begin{equation}\label{SfreeGrav}
    S_\text{free} = -\frac{c^3}{64\pi G}\int d^4x \partial_\alpha h^{ij}\partial^\alpha h_{ij}\,,
\end{equation}
where $d^4x = cdtd^3x$. We point out that since the metric perturbation $\ddot h_{ij}$ in the interaction term is written in the TT gauge, as per Eq.~\eqref{III_Riemann}, $S_{\text{free}}$ must also be written in the TT gauge for consistency. 

In going forward, we make a few simplifying assumptions, following the treatment of~\cite{parikh2020noise,parikh2021signatures}. We orient the coordinates such that the masses are aligned along the $x$-axis and restrict the analysis to the motion along this axis. Furthermore, we consider the interaction of the system only with gravitational waves propagating in the $z$-direction. This setup effectively singles out the interaction of the system with the ``+" polarization mode of the GW since $h_{xx}(z) = h_{+}$. With these considerations, the interaction Lagrangian simplifies to 
\begin{equation}\label{Lint}
L_\text{int} =\frac{m}{4}\ddot{h}_{ij}\xi^i\xi^j=\frac{m}{4}\ddot{h}^{+}(\xi^x)^2.
\end{equation}
The interaction with gravitational modes propagating in the other directions and with the different polarization can also be taken into account, as shown in Appendix~\ref{app:full_pol}. The full treatment does not affect the analysis apart for numerical prefactors of $O(1)$, but is notationally heavy; we thus proceed with the simplified treatment as captured by Eq.~\eqref{Lint}.
With such considerations, the free action for the amplitude $h^+$ in the TT gauge reads 
\begin{align}\label{eq:ActionFreePlus}
     S^{+}_\text{free} = -\frac{c^3}{32\pi G}\int d^4x \partial_\alpha h^+(\textbf{x},t)\partial^\alpha h^+(\textbf{x},t)\,.
\end{align}
   
Finally, we assume that the system, independently of its interaction with gravitational waves, undergoes a motion that can be described in the Fermi frame by a harmonic potential $V = \frac{1}{2} m \omega^2 (\xi^x)^2$ \cite{torovs2024loss} with $\omega\ll\Omega_\text{max}$. Then, adding all the contributions, the full action reads
\begin{align}\label{II:ActionFull}
    S =&\int dt \Big(\frac{m}{2}\dot \xi ^2 - \frac{1}{2}m\omega^2 \xi^2\Big)- \frac{c}{2}\int d^4x \partial_\alpha h(\textbf{x},t)\partial^\alpha h(\textbf{x},t)\nonumber \\
    &+ \int dt\, m\sqrt{\frac{\pi G }{c^2}}\ddot h(0,t)\xi^2\,.
\end{align}
Note that the first term, which refers only to the system, is analogous to that of a single spinless particle in a harmonic potential. In what follows, we will not retain the polarization and the axis label explicitly.  In Eq.~\eqref{II:ActionFull} and in what follows, we use the re-scaled gravitational field $h\rightarrow h\sqrt{16\pi G/c^2} $, to make the orders in the coupling constant $\sqrt{G}$ transparent. 

\section{Challenges posed by standard canonical quantization for open quantum system}\label{sec:oqs}
The first challenge in quantizing the dynamics stems from the fact that the interaction term in the classical Lagrangian, 
\begin{align}\label{eq:L}    
L_{\text{int}} = m\sqrt{\frac{\pi G }{c^2}}\ddot{h}(0,t)\xi^2,
\end{align}    
involves second-order time derivatives of the metric perturbation. The standard, textbook way to derive the Hamiltonian via Legendre transformations works for Lagrangians containing at most first-order time derivatives, while for higher-order derivatives alternative procedures have to be followed \cite{dirac2013lectures, whittaker1964treatise}. In these cases, quantization rests on the use of Ostrogradski's Hamiltonian construction~\cite{whittaker1964treatise}, combined with Dirac theory of constraints~\cite{dirac2013lectures}. 
Although it works for the full system, this procedure is problematic from an open quantum system perspective, since it requires imposing new commutation relations among the conjugate variables, which, in this particular case, imply that the system and environmental  variables do not commute; see, for example, Eq.~\eqref{eq:SENoCommute} in Appendix~\ref{app:DiracQ} and the discussion that follows. Being not decoupled, the set of canonical variables so defined is not adequate for describing an open system dynamics, which requires a clear separation between the system and the environmental degrees of freedom.

One may thus consider that the difficulties with this non-standard quantization procedure can be   overcome by shifting, through an integration by parts, one of the derivatives of $h$ to $\xi$, thus arriving at the classically equivalent Lagrangian $L' = L - dF'/dt$, with  $F' = m\sqrt{\pi G /c^2}\dot h(0,t)\xi^2$. Now the interaction term becomes:
\begin{align}
 L'_\text{int} = -2m\sqrt{\frac{\pi G }{c^2}}\dot h'(0,t)\xi'\dot \xi'\label{L_hxi},
\end{align}
where, for later convenience, we have relabeled the variables as $\xi'$ and $h'$.
The canonical momenta to the variables $\xi'$ and $h'$ can now be computed  using the standard Legendre transformations:
\begin{align}
 p'_\xi &\coloneqq\frac{\partial L'}{\partial \dot\xi'} =  m\dot\xi' -2m\sqrt{\frac{\pi G }{c^2}}\dot h'(0,t)\xi',\label{pxi}\\
p'_h(\textbf{x},t)&\coloneqq \frac{\partial L'}{\partial \dot h'(\textbf{x},t)} = \dot h'(\textbf{x},t) -2m\sqrt{\frac{\pi G }{c^2}}\xi'\dot\xi' \delta^3(\textbf{x})\label{ph}.
\end{align}
However, the canonical variables $(\xi', p'_\xi, h', p'_h)$ defined by $L'$ are still  not suitable to derive the master equation for the reduced density matrix of the system. This is because of the appearance of the environmental degrees of freedom (related to $h'$) in the expression for the conjugate momentum $\hat{p}'_{\xi}$ ascribed to the system. 

We explain the problem by resorting to the toy model  discussed in Ref.~\cite{EqLagrangians2026}. Consider the Lagrangian
\begin{align}\label{eq:toyL'}
L_{\text{toy}}'=\frac{1}{2}m_1 \dot{x}^2_1 + \frac{1}{2}m_2 \dot{x}^2_2-g\dot{x}_1x_2,
\end{align}
which describes the interaction between two distinguishable particles $1$ and $2$. We are interested in the scenario where particle 1 is the system of interest and particle 2 is treated as its environment. 
The main issue  with $L_{\text{toy}}'$, which is the same as that of the Lagrangian $L'$, is that the conjugate momentum it assigns to particle 1, 
\begin{align} \label{eq:p1prime}
p'_1 = \frac{\partial L'_{\text{toy}}}{\partial \dot{x}_1} = m_1\dot{x}_1-g x_2,
\end{align}
does not represent a good observable for particle 1 in an open quantum systems scenario, where particle 2 is traced away and as such is not accessible. This is because, as per Eq.~\eqref{eq:p1prime}, in order to determine $p'_1$, one needs to measure the position $x_2$ of particle 2 in addition to the velocity $\dot{x}_1$ of particle 1. However, and this is where the problem lies,  $x_2$  is supposed not to be accessible since degrees of freedom associated to particle 2 have been traced away.

On a formal level, one can quantize the two-particle system in the standard way starting from $L_{\text{toy}}'$, carry out the calculations and trace over particle 2, eventually arriving at a master equation for particle 1 of the form:
\begin{equation} \label{eq:me_toy}
\partial_t \hat\rho_1 = \mathcal L_{\hat x_1, \hat p_1'}[\hat\rho_1],
\end{equation}
where the super-operator $\mathcal  L_{\hat x_1, \hat p_1'}[\cdot]$ is a function of $\hat x_1$ and  $\hat p_1'$. The master equation is well--defined mathematically, but it is not useful from the physical point of view, being a function of $\hat p_1'$, which cannot be experimentally measured in the open quantum system scenario Eq.~\eqref{eq:me_toy} represents.

In summary, the challenge that one faces in describing open system dynamics starting from the classical action~\eqref{II:ActionFull}, is  that using either the Lagrangian $L$ with the interaction~\eqref{eq:L} or $L'$ with the interaction~\eqref{L_hxi}, the  quantization procedure  does not assign conjugate variables to the system, which are suitable for describing the reduced open quantum system dynamics. 

In the next section we show how these difficulties can be overcome.

\section{Overcoming the challenges via unitary transformation}\label{subsec:UT}
The discussion of the previous section has shown that, in an open quantum system analysis, where the environment eventually is traced out, the starting Lagrangian cannot be chosen arbitrarily within the family of physically equivalent Lagrangians. It should be such that the conjugate variables of the system of interest are operationally defined only in terms of accessible system observables (this is the problem with $L'$) and, when quantized, they should commute with the conjugate variables of the environment (this is the problem with $L$).  Moreover, it would be desirable to work with a Lagrangian which contains at most first-order time derivatives, because in this case the standard quantization procedure can be applied, which is simpler. 

The way through comes from the freedom that quantum mechanics allows in redefining the operators (and consequently the states) through a unitary transformation, without changing the physical predictions. This suggests that we can first quantize the global system and environment starting from Lagrangian $L'$, which is first-order in time. As discussed before, the associated conjugate variables are not suited for performing an open system analysis. To resolve this issue, we then perform a suitable unitary transformation such that the transformed conjugate variables of the system of interest are operationally defined in terms of the system alone, and commute with the transformed conjugate variables of the environment (and vice versa). Such a setup will then be ready to perform a meaningful trace over the environment. 

Let us then quantize the system using the Lagrangian $L'$. The canonical variables $(\xi', p'_\xi, h', p'_h)$ become operators $(\hat\xi', \hat p'_ \xi, \hat h', \hat p'_h)$ satisfying the standard commutation relations:
\begin{align} \label{eq:comm_rel}
[\hat{\xi}',\hat{p}'_{\xi}] = i\hbar,\qquad   [\hat{h}'(\mathbf{x},t),\hat{p}'_{h}(\mathbf{y},t)]  = i\hbar \delta^3(\mathbf{x}-\mathbf{y}),
\end{align}
with all other commutators being zero. The conjugate momenta are operationally defined in Eqs.~\eqref{pxi} and~\eqref{ph}. The next step is to choose a unitary transformation such that the new operator $\hat p_ \xi$, refers only to the observables of the particle and commutes with the gravitational degrees of freedom. This is achieved by the following transformation:
\begin{equation}\label{eq:GlobalUnitary}
\hat{T} = e^{i\hat F/ \hbar}\,, \qquad\hat F=m \sqrt{\frac{\pi G}{c^2}} \hat \xi'{}^{2} \hat p'_h(0,t)\,,
\end{equation}
which relates the primed and unprimed operators as
\begin{align}\label{eq:GlobalUnitary1}
\hat{\xi} &= \hat{T}^{\dagger}\hat \xi'\hat{T}, \qquad \hat{p}_{\xi} = \hat{T}^{\dagger}\hat{p}'_{\xi}\hat{T},\nonumber\\
\hat{p}_{h} &= \hat{T}^{\dagger}\hat{p}'_{h}\hat{T}, \qquad\hat{h} = \hat{T}^{\dagger}\hat h'\hat{T}.
\end{align}
The relationship above can be inverted to express the primed operators in terms of the unprimed ones. One gets:
\begin{align} \label{eq:transf_T}
\hat \xi' &=\hat \xi, \quad \hat p'_{\xi} =\hat p_\xi -2m\sqrt{\frac{\pi G}{c^2}}\hat\xi\hat p_h(0,t), \nonumber \\
\hat p'_h &=\hat p_h\,, \quad\hat h'(\mathbf{x},t) = \hat h(\mathbf{x},t)+m\sqrt{\frac{\pi G}{c^2}}\hat\xi^2\delta^3(\mathbf{x}).
\end{align}
The transformation being unitary, the commutation relations~\eqref{eq:comm_rel} among the operators $(\hat\xi', \hat p'_\xi, \hat h', \hat p'_h)$ remain valid also for the transformed variables $(\hat\xi, \hat p_\xi, \hat h, \hat p_h)$.

Now we show that the new momentum $\hat p_\xi$ has the desired property of referring only to the system observables. We are not attaching any specific physical meaning to it; the point we want to make is that it allows to consistently compute physical quantities related to the system from the master equation, once the gravitational degrees of freedom have been traced out.  For example, consider the velocity $\hat v = \dot{\hat\xi}$ (= $\dot{\hat\xi}'$) of the particle which, according to Eq.~\eqref{pxi}, is related to the momentum $\hat p'_\xi$ as follows:
\begin{equation}
\hat v = \frac{\hat p'_\xi}{m} + 2\sqrt{\frac{\pi G }{c^2}}\dot{\hat h}(0,t)\hat\xi.
\end{equation}
Using Eqs.~\eqref{pxi} and~\eqref{ph}, we can rewrite $\hat v$ in terms of the corresponding canonical variables; to order $O(G)$ it reads
\begin{equation}\label{eq:NonLocalVel2}
\hat v = \frac{\hat p'_\xi}{m} +\frac{2G\Omega_\text{max}^3}{\pi c^5}\{\hat\xi'^2,\hat p'_\xi\} +2\sqrt{\frac{\pi G}{c^2}}\hat p'_h(0,t)\hat \xi'\,,
\end{equation}
where $\Omega^3_{\text{max}}$ appears due to the cutoff imposed in computing $\delta^3(0)$ as detailed in Eq.~\eqref{eq:cutoffDiracDelta}.

We see that it depends also on the environmental degrees of freedom. If now we express the prime variables in terms of the un-primed ones via the transformation in Eq.~\eqref{eq:GlobalUnitary1}, we immediately obtain:
\begin{equation}\label{velocity2}
\hat v =\frac{\hat p_\xi}{m} +\frac{2G\Omega^3_{\mathrm{max}}}{\pi c^5}\{\hat\xi^2,\hat p_\xi\},
\end{equation}
and the dependence on the gravitational degrees of freedom has disappeared. Then, since the master equation for the reduced density matrix $\hat \rho_S(t) = \Tr_E[\hat\rho_{SE}(t)]$ will have the structure $\partial_t \hat\rho_S(t) = \mathcal L_{\hat \xi, \hat p_\xi}[\hat\rho_S(t)]$ (see Sec.~\ref{sec:pme}),  Eq.~\ref{velocity2} allows one to determine $\text{Tr}_S[\hat v^n\hat\rho_S(t)]$, together with the position $\text{Tr}_S[\hat\xi^n \hat\rho_S(t)]$, which, for a single spin-less particle, represent all there is to know. For example, from the velocity, one can compute the kinetic energy, which is given by
\begin{equation}\label{Ekin}
    \hat{\mathscr{E}}_{\text{kin}} =\hat p_\xi^2/2m + \frac{G \Omega^3_{\mathrm{max}}}{\pi c^5}\acomm{\hat p_\xi}{\acomm{\hat p_\xi}{\hat \xi^2}} + \mathcal{O}(G^2)\,. 
\end{equation}
The significance of the relations above is that the position, velocity and conjugate momentum of the particle are related to each other, without making reference to the gravitational degrees of freedom. Also the commutation relations ensure that the two sets of degrees of freedom are not coupled to each other (in the Schr\"odinger picture). We thus see that in the unprimed representation,  all the relevant system observables, which are functions of the position and velocity operator, are described in terms of the system degrees of freedom ($\hat{\xi},\hat{p}_{\xi}$), and can thus be described consistently at the level of the reduced density matrix. 

We note that the requirement that both the proper separation and its velocity be expressible entirely in terms of system canonical variables uniquely fixes the coupling between system and environment in the transformation in Eq.~\eqref{eq:GlobalUnitary}, to order $\mathcal{O}(G)$. More explicitly, the requirement that in going from~\eqref{eq:NonLocalVel2} to~\eqref{velocity2} the term $\propto \hat p'_h(0,t)\hat \xi'$ vanishes uniquely determines the product between system and environment operators chosen in ~\eqref{eq:GlobalUnitary}. Therefore, the transformation is unique up to canonical transformations that are local to the system and environmental sectors, as the latter simply redefine the local variables without changing the system-environment decomposition.

Before closing this section, we point out that as with the environment of gravitational waves,  a similar issue arises in describing open system dynamics for a non-relativistic particle interacting with the environment of electromagnetic field. In  Ref.~\cite{EqLagrangians2026}, it is shown that if the system observables of interest are the position and velocity of the particle, one must work with the conjugate variables obtained from the $\dot{\mathbf{A}}_{\perp}\cdot \mathbf{x}$ interaction Lagrangian, rather than the $\mathbf{A}_{\perp}\cdot \dot{\mathbf{x}}$ one. In fact, the analysis performed in Ref.~\cite{EqLagrangians2026} motivates the choice for the functional form of $\hat{F}$ in Eq.~\eqref{eq:GlobalUnitary}.    

\section{The Hamiltonian}\label{sec:Hamiltonian}
In this section we derive the Hamiltonian $\hat{H}$ starting from $L'$, and express it in terms of the primed and the unprimed conjugate variables. Starting  from $L'$, after a long but straightforward calculation, we get
\begin{align}
     H &= \frac{1}{1-4\pi Gm \xi^2\delta^3(0)/c^2}\times\nonumber\\
     &\Big[\frac{p'_\xi{}^2}{2m} +2\sqrt{\frac{\pi G}{c^2}}p'_h(0,t)p'_\xi\xi' + m\frac{2\pi G}{c^2}p'_h(0,t)^2\xi'^2\Big]\nonumber\\
     &+\frac{1}{2}m\omega^2\xi'^2+\frac{1}{2}\int d^3x \Big(p_h'(\textbf{x},t)^2+c^2\grad h'(\textbf{x},t)^2\Big)\,.
\end{align}
The Hamiltonian appears to be divergent due to the point particle interaction term, which is similar to the divergence that appears in the Hamiltonian of electrodynamics~\cite{gundhi2023motion}.  However, remembering the discussion around Eq.~\eqref{eq:cutoff} of Sec.~\ref{sec:FNC}, that the calculations cannot be trusted on length scales smaller than c/$\Omega_{\text{max}}$, $\delta^3(0)$ must be computed by imposing a UV cutoff inside the integrals, such that $\delta^3(0) =  \Omega^3_{\text{max}}/(\pi^2 c^3)$ remains finite; see Appendix~\ref{App_Reg}. Further, since we are ultimately interested in deriving the master equation for the matter system to leading order in the gravitational coupling, we consistently truncate the Hamiltonian at second order in $\sqrt{G}$. Quantization is implemented by imposing the canonical commutation relations and by adopting Weyl's operator ordering. Following this procedure, we arrive at the Hamiltonian operator
\begin{align} \label{eq:H'}
     \hat H = &\frac{\hat p'_{\xi}{}^2}{2m} +\frac{1}{2}m\omega^2\hat \xi'{}^2+\frac{ G\Omega^3_{\mathrm{max}}}{\pi c^5}\acomm{\hat\xi'{}^2}{ \hat p'_{\xi}{}^2}\nonumber\\
     &+\sqrt{\frac{\pi G}{c^2}}\hat p'_{h}(0,t)\acomm{\hat p'_{\xi}}{\hat\xi'}+2\pi\frac{mG}{c^2}(p'_h(0,t))^2\xi'{}^2\nonumber\\
     &+\frac{1}{2}\int d^3x \Big(\hat p'_{h}(\textbf{x},t)^2+c^2\grad \hat h'(\textbf{x},t)^2\Big)\,,
 \end{align}
where $\{\hat{A},\hat{B}\} := \hat A\hat B+\hat B\hat A$.  Now we change the variables from the primed to the un-primed ones. Using Eq.~\eqref{eq:transf_T} to second order in $\sqrt{G}$, we get (Appendix~\ref{app_pwz})
\begin{align}\label{FH_FinalH}
    &\hat H = \nonumber\\
    &\frac{\hat p_\xi^2}{2m} +\frac{G\Omega^3_{\mathrm{max}}}{\pi c^5}\acomm{\hat \xi^2}{\hat p_\xi^2}+\frac{1}{2}m\omega^2\hat\xi^2 -\frac{\pi G m^2}{2}\hat \xi^4\grad^2\delta^3(0)\nonumber\\
    &-\sqrt{\frac{\pi G}{c^2}}mc^2\nabla^2\hat h(\textbf{x},t)\big|_{\mathbf{x}=0}\hat \xi^2\nonumber\\
    &+\frac{1}{2}\int d^3x \Big[\hat p_h(\textbf{x},t)^2+c^2\nabla \hat h(\textbf{x},t)^2\Big]\nonumber\\
    &\coloneqq \hat H_{\text{sys}}+ \hat H_\text{int} + \hat H_E  -\frac{\pi G m^2}{2}\hat \xi^4\grad^2\delta^3(0)\,.
\end{align}
The first line of the full Hamiltonian~\eqref{FH_FinalH} is comprised only of the system degrees of freedom, and reduces to the sum of potential and kinetic energies in the absence of interaction (the limit $G\to0$). It thus motivates the definition
\begin{align}\label{eq:Hsys}
\hat{H}_{\text{sys}} = \frac{\hat p_\xi^2}{2m} +\frac{1}{2}m\omega^2\hat\xi^2+\frac{G\Omega^3_{\mathrm{max}}}{\pi c^5}\acomm{\hat \xi^2}{\hat p_\xi^2}\,,
\end{align}
which governs the system's evolution. Note that we have isolated the last term in the first line $(\grad^2\delta^3(0)\propto \Omega^5_{\text{max}})$, since it will be canceled exactly at the level of the reduced density matrix, by an additional contribution coming from the system-environment interaction (c.f. Appendix~\ref{APP_markovLim}). The second line of the Hamiltonian~\eqref{FH_FinalH} is the interaction Hamiltonian $\hat{H}_{\text{int}}$, where we see that the gravitational field couples to the proper relative distance $\hat\xi$ or, more precisely, to the mass quadrupole $\hat I = m\hat\xi^2$ of the system of two masses considered in this work. Finally, the last line of Eq.~\eqref{FH_FinalH} is the Hamiltonian referring to the environmental degrees of freedom, which reduces to the Hamiltonian of the freely propagating gravitational waves, in the absence of interactions, as per Eq.~\eqref{eq:transf_T}.

Equation~\eqref{FH_FinalH} is the first key result of the present work: together with the variables $(\hat \xi, \hat p_\xi, \hat h, \hat p_h)$, it represents the proper formulation of the Hamiltonian of the system, that allows a consistent open quantum system analysis, yielding a master equation defined in terms of physically accessible observables of the system of interest. We will derive this master equation in Sec.~\ref{sec:pme} .

The Hamiltonian we derived in Eq.~\eqref{FH_FinalH}, to describe open dynamics of a system interacting with weak gravity, differs from the previous ones used in the literature \cite{malcolm2025signatures, cho2025non, sharifian2024open,torovs2024loss, parikh2021signatures, parikh2020noise, blencowe2013effective,anastopoulos2013master, breuer2009metric,oniga2016,oniga2017quantum,asprea2021gravitational,asprea2021letter,anastopoulos1996quantum}. 

For instance,  Refs.~\cite{malcolm2025signatures, parikh2021signatures,parikh2020noise} work with the Hamiltonian~\eqref{eq:H'}, using which the Langevin equation for $\xi$ is derived in Refs.~\cite{parikh2021signatures,parikh2020noise}, while the master equation is derived in \cite{malcolm2025signatures}. Since in this work our aim is to obtain the latter, we emphasize that the conjugate variables assigned by the Hamiltonian~\eqref{eq:H'} to the system and the environment, do not enable the description of the system observables, such as the velocity operator, at the level of the reduced density matrix. It thus motivates the need for re-parameterizing the Hamiltonian for it to be suitable for open system dynamics. We achieve this in Eq.~\eqref{FH_FinalH}. 

Refs.~\cite{cho2025non, torovs2024loss,sharifian2024open}, on the other hand, start with the interaction Lagrangian~\eqref{eq:L}. However, these works deal with second order time derivative in Eq.~\eqref{Lint} by ignoring the interaction term altogether for defining the conjugate variables. Thus, they obtain the free Hamiltonian $\hat{H}_{\text{free}}$ from the free Lagrangian following standard Legendre transformation, and \textit{assume} 
\begin{align}
\hat{H}_{\text{int}} =-L_{\text{int}}(\hat{h},\hat{\xi}) =-  m\sqrt{\frac{\pi G }{c^2}}\ddot{\hat{h}}(0,t)\hat{\xi}^2.
\end{align}
However, this prescription is not entirely satisfactory, and must be viewed as a further approximation, since the definition of conjugate variables depends upon the interaction Lagrangian, as per the Dirac prescription. See, for example, Appendix~\ref{app:DiracQ}. Since the quantization scheme used by the authors differs from our derivation, they arrive at a different final Hamiltonian compared to ours in \eqref{FH_FinalH}.  

Lastly, older literature \cite{blencowe2013effective,anastopoulos2013master, breuer2009metric,oniga2016,oniga2017quantum,asprea2021gravitational,asprea2021letter,anastopoulos1996quantum} formulates the dynamics in Cartesian coordinates rather than working with the Fermi coordinates. A comparison between these two approaches is already performed in Ref.~\cite{malcolm2025signatures}. As we already pointed out in Sec.~\ref{sec:FNC}, we consider Fermi coordinates to be more appropriate in order to work with physical observables.

\section{The master equation}\label{sec:pme}
The master equation is obtained by tracing over the conjugate variables assigned to the linearized gravitational field. Here we briefly summarize the derivation, which follows the standard Feynman--Vernon treatment of open quantum systems~\cite{calzetta2009nonequilibrium}. We take as our starting point the full Hamiltonian in Eq.~\eqref{FH_FinalH} and assume, as customary for open quantum systems~\cite{breuer2002theory,schlosshauer2007decoherence}, an initially factorized system--environment state, $\hat\rho(0)=\hat\rho_S\otimes\hat\rho_E$, with $\hat\rho_E$ assumed to be a Gaussian state with vanishing average. Then, the evolution of the reduced density matrix is expressed as a closed-time-path functional integral over two copies of the system trajectory~\cite{calzetta2009nonequilibrium}. See also Ref.~\cite{gundhi2023motion}, where the derivation of the master equation is presented using the influence functional formalism, for a non-relativistic charged particle interacting with the electromagnetic field.

The gravitational variables enter quadratically in the environmental action and linearly in the interaction Hamiltonian, and can therefore be integrated out explicitly. Because the environmental state is Gaussian and has zero mean, the first-order contribution in the interaction vanishes, while the influence functional at order \(\mathcal O(G)\), is entirely determined by the two-point correlation function of the gravitational field.
Differentiating the reduced propagator with respect to time yields the non-Markovian master equation 

\begin{align}\label{ME_nonMarkov}
    \partial_t \hat\rho_t =& -\frac{i}{\hbar}\comm{\hat H_{\text{sys}} -\frac{\pi G m^2}{2}\hat \xi^4\grad^2\delta^3(0)}{\hat\rho_t}\nonumber \\
    &+\frac{i}{2\hbar}\int_0^t d\tau\mathcal{D}(\tau)\comm{\hat\xi^2}{\acomm{\hat{\xi}_{H_0}^2(-\tau)}{\hat\rho_t}}\nonumber\\
    &-\frac{1}{\hbar}\int_0^td\tau\mathcal{N}(\tau)\comm{\hat\xi^2}{\comm{\hat{\xi}_{H_0}^2(-\tau)}{\hat\rho_t}}\,,
\end{align}
where $\hat \xi_{H_0}^2(-\tau)$ is the Heisenberg picture operator evolved using the system Hamiltonian $\hat{H}_{\text{sys}}$ in the absence of gravitational interaction, i.e., with $\hat{H}_{0} = \hat{H}_{\text{sys}}|_{G=0}$. Since the derivation of the master equation follows closely the derivation presented in Ref.~\cite{gundhi2023motion} in the context of electrodynamics, the functional form of the master equation for linearized gravity can be obtained by replacing $\hat{x}$ and $\hat{x}_{\text{\tiny{H}}_{s}} (-\tau)$ in Eq.~(\!(42)\!) of Ref.~\cite{gundhi2023motion} with $\hat{\xi}^2$ and $\hat{\xi}^2_{H_0}(-\tau)$ respectively, and similarly by replacing the system Hamiltonian for electrodynamics $\hat{H}_s$ in Eq.~(\!(42)\!) of \cite{gundhi2023motion} with the first line of the full Hamiltonian in Eq.~\eqref{FH_FinalH}. 
The kernels $\mathcal{N}$ and $\mathcal{D}$ are also defined analogously, as in Eq.~(\!(33)\!) in Ref.~\cite{gundhi2023motion}, in terms of the two-time correlation of the environmental operator coupled to the system~\cite{calzetta2009nonequilibrium}. For linearized gravity they read
\begin{align}
&\mathcal{N}(\tau) =\nonumber\\
&\frac{m^2\pi G}{2\hbar c^2}\langle\Big\{c^2\nabla^2\hat h(\textbf{x},t+ \tau)|_{x=0},c^2\nabla^2\hat h(\textbf{x},t)|_{x=0}\Big\}\rangle_{\hat\rho_E}\label{noiseKer}\,,\\
&\mathcal{D}(\tau)=\nonumber\\ 
&\frac{i}{\hbar}\frac{m^2\pi G}{c^2}\langle\Big[c^2\nabla^2\hat h(\textbf{x},t+ \tau)|_{x=0},c^2\nabla^2\hat h(\textbf{x},t)|_{x=0}\Big]\rangle_{\hat\rho_E}\label{dissKer}\,.
\end{align}
Thus, in terms of the kernels $\mathcal{N}$ and $\mathcal{D}$, we see that the first line of the master equation~\eqref{ME_nonMarkov} represents the  von Neumann evolution, while the second and the third lines capture dissipation and decoherence respectively, which are quantitatively described by the kernels $\mathcal{D}$ and $\mathcal{N}$.

We now specialize the master equation~\eqref{ME_nonMarkov} to the case where the initial state of the gravitons is assumed to have a thermal distribution, i.e., $\hat \rho_E = e^{-\hat H_E/k_bT}/Z$, where $Z$ is the partition function.  The case study for the thermal state should be understood as a reference which allows for a quantitative and a concrete analysis of the master equation, capturing the essential physics. This choice does not encompass all physically motivated graviton states: indeed, primordial gravitational fluctuations are generally expected to be squeezed~\cite{Grishchuk1989QuantumState, Martin2008}, encompassing correlations absent in a thermal state. Nonetheless, the thermal calculation serves as a useful benchmark to analyze the main properties of the master equation~\eqref{ME_nonMarkov}.

We will consider the long time limit such that $t \gg 1/\Omega_\text{max}$ and  $t\gg \hbar/(k_{b}T)$ where, as shown in Appendix~\ref{App_ker}, the noise and dissipation kernels can be computed explicitly and acquire a simple mathematical form:
\begin{align}\label{PME_kerDelta}
    \mathcal{N}(\tau) &= \frac{1}{2}\frac{m^2 G}{c^5}\frac{k_b T}{\hbar}\frac{d^4}{d\tau^4}\delta(\tau)+ \mathcal{N}_0(\tau),\nonumber\\
    \mathcal{D}(\tau) &= -\frac{\theta(\tau)}{2}\frac{m^2 G}{c^5}\frac{d^5}{d\tau^5}\delta(\tau),
\end{align}
with $\theta(\tau)$ being the Heaviside theta function. Note that in the expression for $\mathcal{N}$ we have explicitly separated the thermal contributions to the vacuum one $\mathcal{N}_0$, which reads
\begin{align}\label{NoiseKerVac}
    \mathcal{N}_0(\tau)= \frac{30 m^2G}{\pi c^5} \frac{\epsilon^6 - 15 \epsilon^4 \tau^2 + 
   15 \epsilon^2 \tau^4 - \tau^6}{(\epsilon^2 + \tau^2)^6}\,.
\end{align}
By inserting the expressions of the kernels in Eq.~\eqref{PME_kerDelta} into the master equation~\eqref{ME_nonMarkov}, we arrive (see Appendix~\ref{APP_markovLim}) at the following expression of the master equation in the long-time regime:
\begin{widetext}
\begin{align}\label{PME_MEFull}
    \partial_t\hat \rho_t = -&\frac{i}{\hbar}\comm{ \hat{H}_{\text{sys}}-\frac{G m^2\omega^2 g(\omega,\Omega_{\text{max}})}{\pi c^5}\hat \xi^4}{\hat\rho_t}\nonumber\\
    -&\frac{i}{\hbar}\frac{2mG\omega^4}{ c^5}\comm{\hat\xi^2}{\acomm{\acomm{\hat\xi}{\hat p_\xi}}{\hat\rho_t}}-\frac{i}{\hbar}\frac{G g(\omega,\Omega_{\text{max}})}{\pi c^5}\comm{\hat\xi^2}{\acomm{\hat p_\xi^2}{\hat\rho_t}}\nonumber\\
   -&\frac{2m^2G\omega^5}{\hbar c^5}\Big(1+\frac{k_bT}{\hbar\omega}\Big)\comm{\hat \xi^2}{\comm{\hat \xi^2}{\hat \rho_t}} + \frac{2G\omega^3}{\hbar c^5}\Big(1+\frac{k_bT}{\hbar\omega}\Big)\comm{\hat \xi^2}{\comm{\hat p_\xi^2}{\hat \rho_t}}- \frac{1}{\hbar}\frac{m G}{\pi c^5}f(\Omega_\text{max},\omega,T)\comm{\hat\xi^2}{\comm{\acomm{\hat\xi}{\hat p_\xi}}{\hat\rho_t}},
\end{align}
\end{widetext}
where
\begin{align}
g :=& \Omega_{\mathrm{max}}(\Omega_\text{max}^2+ 2\omega^2)\,,\nonumber\\    
f :=& \frac{3}{2}\Omega_\text{max}^4+ \frac{k_bT}{\hbar}\Omega_\text{max}^3+2\frac{k_bT}{\hbar}\Omega_\text{max}\omega^2+\Omega_\text{max}^2\omega^2 +\nonumber\\
    &-4\omega^4\log\Big(\frac{2\omega}{\Omega_\text{max}}\Big) -4\gamma_{EM}\omega^4\,,
\end{align}
and $\gamma_\text{EM}$ is the Euler-Mascheroni constant. The master equation ~\eqref{PME_MEFull} is one of the main results of this work. 

We note that there are many contributions proportional to the cutoff frequency $\Omega_\text{max}$ that appear in the master equation. While this might look problematic, as it suggests the presence of divergences, we emphasize that this is not the case, in the regime of validity of the approximations used to derive the master equation. In particular,  $\Omega_{\text{max}}$ cannot take arbitrarily large values, since otherwise it would not respect the second order expansion of the metric in Eq.~\eqref{III_TTmetric} as we elaborated in Sec.~\ref{sec:FNC}.  

As we already mentioned earlier the term $\frac{\pi G m^2}{2}\hat \xi^4\grad^2\delta^3(0)$, present in the full Hamiltonian~\eqref{FH_FinalH}, does not appear since it is canceled by an additional term which comes from the integral involving $\mathcal{D}(\tau)$. Furthermore, the other two cutoff dependent terms in the first line--one proportional to $g(\omega,\Omega_{\text{max}})$ (also coming from the integral involving the dissipation kernel),  and the other one present in $\hat{H}_{\text{sys}}$ specified in Eq.~\eqref{eq:Hsys}---when treated as c-numbers (to give an order of magnitude estimate), lead to an effective Hamiltonian of the form $\frac{p^2}{2m}(1+\Delta_{p_\xi})+\frac{1}{2}m\omega^2\xi^2(1-\Delta_\xi)$ with
\begin{equation}
    \Delta_{p_\xi}\sim \Delta_{\xi} \sim \frac{m}{m_P}\frac{l_P}{\lambda_\text{min}}\Big(\frac{\xi}{\lambda_\text{min}}\Big)^2\,.
\end{equation}
Here, we have neglected factors of $\mathcal{O}(1)$ and we have introduced the Planck mass $m_P = \sqrt{\hbar c/G}$, Planck length $l_P = \sqrt{\hbar G/c^3}$, and $\lambda_\text{min} = 2\pi c/\Omega_\text{max}$ the minimum wavelength mode of the gravitational waves. We then see that consistency with the constraints $\lambda_{\text{min}}\gg l_P$ imposed by quantum gravity considerations, and $\lambda_{\text{min}}\gg \xi$ imposed by  Eq.~\eqref{eq:cutoff}, render the $\Omega_{\text{max}}$ dependent corrections in the unitary part of the dynamics to besmall, unless $m\gg m_P$.
However, using $\lambda_\text{min}\gg \xi$, we have
\begin{equation}\label{deltasCutoff}
    \Delta_{p_\xi}\sim \Delta _\xi \ll \frac{Gm}{c^2\xi}\,.
\end{equation}
Hence, as long as the system has sufficiently small gravitational compactness, i.e. $Gm/c^2\xi \ll 1$, such contributions will be subdominant. This in fact will be the case for a typical laboratory setting: for example, considering the setup in Ref.~\cite{deplano2024coulomb} consisting of two $m\sim 10^{-18}$kg nanoparticles separated by $\xi\sim 10^{-5}$m, Eq.~\eqref{deltasCutoff} results in $\Delta_\xi, \Delta p_\xi\ll 10^{-40} $. A summary of the conditions which ought to be satisfied for the master equation~\eqref{PME_MEFull} to be valid is presented in Table~\ref{tab:validity}.
\begin{table}[t]
\centering
\begin{tabular}{c|c}
\hline
Approximation & Regime of validity \\
\hline\hline
Long-wavelength 
& $\xi\ll \lambda_{\min}$ \\[4pt]
$\Delta_{\xi}\ll 1,\Delta_{p_\xi}\ll 1$
& $Gm/c^2\xi\ll 1$ \\[4pt]
Markovian regime
& $t \gg \omega^{-1}$\\
\hline
\end{tabular}
\caption{Summary of the approximations adopted to derive the master equation Eq.~\eqref{PME_MEFull} and their regime of validity . %Conditions for the applicability of the master equation. Here, $\xi$ is the average relative separation and
}
\label{tab:validity}
\end{table}

As a consequence of the non--linear system--environment coupling, the master equation for linearized gravity differs in its functional form from the well-known phenomenological master equations, such as the Caldeira-Leggett equation. The physical contents of the non-unitary parts of the master equation~\eqref{PME_MEFull} are thus not immediately transparent. Although a more tractable limit will be investigated in the next section, several physical conclusions can already be drawn from Eq.~\eqref{PME_MEFull}. Indeed, the second line of the master equation~\eqref{PME_MEFull}, which derives from the dissipation kernel, describes the energy lost by the system to the environment. As detailed in appendices~\ref{app:GravBremsstrahlung} and~\ref{App_EOMs}, if the system is in a semiclassical state, e.g.,~a highly populated coherent state with an initial amplitude $\langle\hat \xi\rangle = l_0$, it predicts that, averaging over many periods $T$, the system loses energy according to 
\begin{align}\label{powerloss_main}
    \langle d\langle \hat{\mathscr{E}}\rangle/dt\rangle_T &= -\frac{16}{15}\frac{G I^2\omega^6}{c^5}\,, 
\end{align}
where $I= m l_0^2$ and $\langle\cdot\rangle_T$ denotes the time average. We see that Eq.~\eqref{powerloss_main} consistently matches the classical formula of energy loss by a system emitting gravitational waves~\cite{maggiore2008gravitational}. 
The last line of~\eqref{PME_MEFull} collects all the terms coming from the noise kernel, and describes the fluctuations induced by environmental noise, in particular decoherence. While the first and third term can be considered as analogous to, respectively, the normal and the so-called anomalous diffusion in (linear) quantum brownian motion (QBM)~\cite{hu1992quantum,caldeira1983path}, the second term arises entirely due to the non-linearity of the system-environment coupling in $\hat H_\text{int}$~\cite{hu1993quantum}. As shown in Appendix~\ref{app:GravBremsstrahlung}, the interplay between this fluctuating backaction and dissipation approximately drives the system toward a stationary state determined by the bath temperature, in agreement with the expected thermodynamic behavior.
Finally, we highlight that the master equation~\ref{PME_MEFull}, which is obtained to second order in $\sqrt{G}$, and within the Markovian/late-time limit, is not of the Lindblad form and hence does not describe a completely positive evolution~\cite{gorini1976completely,lindblad1976generators,benatti2005open}. However, this does not imply that the underlying dynamics is unphysical since the exact reduced evolution from separable initial conditions, as is the case here, is always completely positive~\cite{breuer2002theory}. %Therefore the lack of complete positivity arises from the approximations such approximation s may exhibit positivity violations for certain initial states~\cite{benatti2005open}.
Therefore, the lack of complete positivity arises from the approximations employed; nonetheless, within the regime of validity of such approximations, master equations like~\eqref{PME_MEFull} are expected to provide a meaningful perturbative description of the underlying completely positive reduced dynamics~\cite{hartmann2020accuracy}, as supported by the semiclassical calculations in Appendix~\ref{app:GravBremsstrahlung}.

\section{Resemblance to the Caldeira-Leggett master equation for small fluctuations of the system}\label{Sec:lin} 

Until now we have been treating gravity as a classical (Minkowski) background with a quantized perturbation $\hat h$ on top, while matter has been treated fully quantum-mechanically. In this section we further explore the relevant scenario where matter too can be described by an effectively classical background with small fluctuations, quantizing only the latter. We thus start by considering the split
\begin{align}
\xi \rightarrow \xi_{cl} +\delta \xi
\end{align} 
in the action~\eqref{II:ActionFull} (and, as before, $g_{\mu\nu} \rightarrow \eta_{\mu\nu}+ h_{\mu\nu}$) and retain terms up to second order in the perturbations. Here, $\xi_{cl}$ is the solution to the Euler-Lagrange equation of motion for the background dynamics, treated classically. We describe the classical motion via the Lagrangian $L_\text{cl} =m\dot \xi_{cl} ^2/2 - m\omega^2 (\xi_{cl}-l_0)^2/2$, where we have modified the harmonic potential to be centered at an average distance $l_0$. We do so in order to consider the solution of the classical equations of motion given by $\xi_{cl}(t) = l_0$, which means that in absence of gravitational waves, the two masses are at rest at a fixed distance, and, thus, $\delta\xi$ are perturbations of this configuration.

With these considerations, the action for the fluctuations reads
\begin{align}
     S_{\text{fluc}} =&-\frac{c}{2}\int d^4x \partial_\alpha h(\textbf{x},t)\partial^\alpha h(\textbf{x},t)\nonumber\\ 
     &+\int dt \Big(\frac{m}{2}\dot \delta\xi ^2 - \frac{1}{2}m\omega^2 \delta\xi^2\Big)\nonumber\\
    &- 2\int dt m\sqrt{\frac{\pi G}{c^2}}\dot h(0,t)\frac{d}{dt}\big(\xi_{cl}\delta \xi\big)\,.
\end{align}
Just like in the full quantum mechanical treatment of matter, because of the derivative coupling, a standard Legendre transformation defines canonical variables that are not suitable for open quantum system analysis for the perturbations. Therefore, to overcome this issue, as before, we perform a global unitary transformation defined by $\hat T_\text{lin} = e^{i\hat F_\text{lin}/\hbar}$, with $\hat F_\text{lin} = 2ml_0\sqrt{\pi G /c^2}\hat p_h(0,t)\hat{\delta\xi}$, giving the Hamiltonian
\begin{align}
    \hat{H} =& \frac{\hat{\delta p_\xi}^2}{2m}\Big(1+\frac{4\pi Gm l_0^2}{c^2}\delta^3(0)\Big)-2ml_0\sqrt{\frac{\pi G}{c^2}}(c^2\grad^2\hat h)|_0\hat{\delta\xi} \nonumber\\
    &+\frac{1}{2}m\hat{\delta\xi}^2\Big(\omega^2-4\pi mGl_0^2\grad^2\delta^3(\textbf{x})|_0\Big)\nonumber\\
    &+\frac{1}{2}\int d^3x \Big(p_h(\textbf{x},t)^2+c^2\grad h(\textbf{x},t)^2\Big)\,.
\end{align}
The first and third terms are, respectively, the kinetic and potential energy of the matter fluctuation $\delta\xi$ where we see that the cutoff dependent terms that were present in Eq.~\eqref{FH_FinalH} reduce to a modification of the mass and frequency parameters. The second term is the leading linear coupling between the matter fluctuation and the gravitational perturbation with the coupling constant now proportional to the average distance $l_0$. Finally, the last term is the free Hamiltonian of the gravitational perturbation.

Then, following the same procedure as in Sec.~\ref{sec:pme}, we can derive the master equation with the linear interaction Hamiltonian $\hat H_\text{int,\ lin} = - 2 ml_0\sqrt{\frac{\pi G}{c^2}}c^2\nabla^2\hat h\big|_0\hat{\delta \xi}$.  Since the envirnmental operator coupled to the system in $\hat H_\text{int,\ lin}$ has not changed, the gravitational kernels will remain the same as those in~\eqref{noiseKer}-\eqref{dissKer}. Thus, the master equation for $\hat{\delta \xi}$ takes the form
\begin{align}\label{eq:masterEqLinear}
   \partial_t \hat\rho_t =& -\frac{i}{\hbar}\comm{\hat H_{\text{sys}}-\hat{\delta\xi}^2\Big(2\pi m^2Gl_0^2\grad^2\delta^3(\textbf{x})|_0\Big) }{\hat\rho_t}\nonumber\\
    &+i\frac{2l_0^2}{\hbar}\int_0^t d\tau\mathcal{D}(\tau)\comm{\hat{\delta\xi}}{\acomm{\hat{\delta\xi}_{\hat H_0}(-\tau)}{\hat\rho_t}}\nonumber\\
    &-\frac{4l_0^2}{\hbar}\int_0^td\tau\mathcal{N}(\tau)\comm{\hat{\delta\xi}}{\comm{\hat{\delta\xi}_{\hat H_0}(-\tau)}{\hat\rho_t}}\,,
\end{align}
where 
\begin{align}
    \hat{H}_{\text{sys}} =& \frac{\hat{\delta p_\xi}^2}{2m}+\frac{1}{2}m\hat{\delta\xi}^2\omega^2 +\frac{2Gl^2_0\Omega^3_{\mathrm{max}}}{\pi c^5}\hat{\delta p}_\xi^2,
\end{align}
and 
\begin{align}
\hat{\delta\xi}_{\hat H_0}(-\tau)=\hat{\delta\xi}\cos \omega \tau -\hat{\delta p_\xi}\sin(\omega \tau)/m\omega.
\end{align} 

We note that the structure of the master equation~\eqref{eq:masterEqLinear}  for the system-perturbation can also be derived after setting $\hat{\xi}\rightarrow \xi_{\text{cl}}\hat{I}+\hat{\delta\xi}$, and the system potential to $m\omega^2\hat{\xi}^2/2\rightarrow m\omega^2(\hat{\xi}-l_0\hat{I})^2/2$ in the general master equation~\eqref{ME_nonMarkov}. In this context, the classical background solution $\xi_{\text{cl}}$ is understood to be the expectation value of the (quantum) system, as described by free evolution in Eq.~\eqref{ME_nonMarkov}, such that  $\xi_{cl} = \langle\hat{\xi}_{\hat{H}_0}(t)\rangle = l_0$. However, in the light of the non-linear system-environment interaction, the proper procedure is to linearize the system perturbation from the outset, at the level of the Lagrangian itself, and then derive the corresponding master equation.

Using the same noise and dissipation kernels as before, i.e., assuming the gravitational field to be initially in a thermal state, and computing the integrals involving the kernels in the same way we did for the general master equation, we obtain the following master equation for the perturbations
\begin{align}\label{App_lin:MElin}
    \partial_t \hat \rho_t= & -\frac{i}{\hbar}\comm{\frac{\hat{\delta p_\xi}^2}{2m_R}+\frac{1}{2}m_R\omega_R^2\hat{\delta\xi}^2}{\hat\rho_t}\nonumber\\
    &-\frac{4ml_0^2G}{\pi\hbar c^5}f_{\text{lin}}(\Omega_\text{max},\omega, T)\comm{\hat{\delta \xi}}{\comm{\hat{\delta p_\xi}}{\hat\rho_t}}\nonumber\\
    &-\frac{i}{\hbar}\frac{l_0^2Gm \omega^4}{2c^5}\comm{\hat{\delta\xi}}{\acomm{\hat{\delta p_\xi}}{\hat \rho_t}}\nonumber\\
    &-\frac{1}{\hbar}\frac{l_0^2Gm^2\omega^5}{c^5}\Big(\frac{1}{2}+\frac{k_bT}{\hbar \omega}\Big)\comm{\hat{\delta\xi}}{\comm{\hat{\delta \xi}}{\hat\rho_t}}, 
\end{align}
where we have defined the following constants
\begin{align}
    m_R &=  m\Big(1-\frac{4}{\pi}\frac{mG}{c^5}\Omega_\text{max}^3l_0^2\Big),\\
    \omega_R^2 &=\omega^2\Big(1-\frac{2}{\pi}\frac{mG}{c^5}\omega^2\Omega_\text{max} l_0^2\Big),\\
    f_\text{lin} &= f(\Omega_\text{max},\omega/2, T).
\end{align}
As before, the term proportional to $\Omega_{\text{max}}^5$ present in the first line of the master Eq.~\eqref{App_lin:MElin} is canceled by an additional term which appears when the integral involving $\mathcal{D}(\tau)$ is computed, and is thus not relevant for the effective system dynamics. 
We recognize that the master equation~\eqref{App_lin:MElin} has the familiar of a linear QBM master equation~\cite{hu1992quantum,ford2001exact}, for a supra-ohmic environmental spectrum $\propto \omega^5$.

\section{Decoherence}\label{sec:deco}
We now turn to the decoherence predicted by the two master equations derived in the previous sections. Let us consider first Eq.~\eqref{PME_MEFull} for which the relevant contributions for decoherence are given by the last line. While the first of these terms is the non-linear analogue of the standard decoherence term in  QBM, the remaining noise-induced contributions contain products of $\hat p_\xi$ and $\xi$, which makes the exact treatment of decoherence more involved, for instance, compared to QBM. Nevertheless, for the purpose of extracting a conservative estimate of the decoherence rate, we restrict attention to the contribution proportional to the double commutator in $\hat\xi^2$. Neglecting vacuum fluctuations, it reads 
\begin{equation}\label{deco_DecoXiXi}
    \partial_t\hat\rho_t = -\frac{2m^2G\omega^4}{c^5\hbar}\frac{k_bT}{\hbar}\comm{\hat\xi^2}{\comm{\hat\xi^2}{\hat\rho_t}}.
    %+\frac{2G\omega^2}{c^5\hbar^2\beta}\comm{\hat\xi^2}{\comm{\hat p_\xi^2}{\hat\rho(t)}}.
\end{equation}
This contribution to decoherence can be evaluated in the position basis giving $\rho_t(\xi,\xi') = e^{-\Gamma t}\rho_0(\xi,\xi')$, where $\rho_t(\xi,\xi') = e^{-\Gamma t}\rho_0(\xi,\xi')$ and 
\begin{align}\label{cohDecay}
    \Gamma
    = \frac{2\omega^4\Delta I^2}{E_P^2}\frac{k_bT}{\hbar}, \qquad \Delta I = m(\xi^2-\xi'^2).
\end{align}
Here, we have identified the mass quadrupole of the system $I = m\xi^2$. We see therefore an exponential decay of coherence between states with different proper distances $\xi$. However, differently from standard mechanisms for decoherence in position~\cite{joos1985emergence,gallis1990environmental, breuer2002theory}, the quadratic dependence on $\hat \xi$ in Eq.~\eqref{deco_DecoXiXi} implies that configurations with exactly equal and opposite relative displacements ($\xi = -\xi'$) are robust to the leading order decoherence effects. For example, we can consider a test mass prepared in a superposition of two spatially localized states on opposite sides of the reference mass, with relative coordinates $\xi=\pm a$ along the same axis. Although spatially distinct, these configurations have the same mass quadrupole and are therefore indistinguishable to the gravitational environment. Notice that, substituting the harmonic potential $V = m\omega^2\xi^2/2$ in Eq.~\eqref{cohDecay}, the rate of decoherence can be rewritten as
\begin{equation}\label{VI_rate}
    \Gamma = 8\Big(\frac{\Delta V}{E_P}\Big)^2\frac{k_b T}{\hbar},
\end{equation}
which depends explicitly on the ratio of the difference in potential energies of the system in the two configurations and the Planck energy. While estimates similar to Eqs.~\eqref{cohDecay} and~\eqref{VI_rate} have also been obtained in previous works~\cite{kanno2021noise,cho2022quantum,moreira2025graviton,torovs2024loss,sharifian2024open,cho2025non,cho2023graviton}, our results contrast with those obtained in Refs.~\cite{anastopoulos1996quantum,blencowe2013effective,anastopoulos2013master,asprea2021gravitational,asprea2021letter,breuer2009metric}, which arrive at a different expression for the decoherence rate $\Gamma$, and predict the loss of coherence to scale with different values of the kinetic energies $ p^2/2m$ corresponding to different superposed configurations, rather than the differences in potential energies as per Eqs.~\eqref{cohDecay} and~\eqref{VI_rate}.

The mechanism driving decoherence is the loss of which-path information to the environment in the form of gravitational bremsstrahlung. Notice that as $\omega \rightarrow 0$, the decoherence rate of Eq.~\eqref{cohDecay} goes to zero to leading order in $G$. One might find this result to be counterintuitive, since even in the absence of bremsstrahlung, there can be some decoherence due to graviton scattering. However, decoherence due to scattering depends on the square of S-matrix elements~\cite{gallis1990environmental,joos1985emergence} and would thus show up at orders $\mathcal{O}(G^2)$ or higher~\cite{donoghue2017epfl} in the master equation. Since the master equation that we derive is to order $\mathcal{O}(G)$, it is consistent to not find decoherence due to scattering in our results. Decoherence due to graviton scattering  thus lies beyond the perturbative regime considered here.

It is also interesting to calculate the decoherence predicted by the master equation~\eqref{App_lin:MElin}. Just as the non-linear case, the master equation~\eqref{App_lin:MElin} describes decoherence in the position basis. Isolating contribution analogous to Eq.~\eqref{deco_DecoXiXi}, and neglecting the vacuum contribution, we find $ \rho_t(\delta\xi,\delta\xi') = e^{-\Gamma t}\rho_0(\delta\xi,\delta\xi'),$ where
\begin{equation}\label{Deco_Lin}
    \Gamma_\text{lin} = \frac{(m\omega^2l_0^2)m\omega^2(\delta\xi-\delta\xi')^2}{E_P^2}\frac{k_bT}{\hbar}\,.
\end{equation}
As previously, $\Gamma_\text{lin}$ is determined by the ratio of the relevant energy scales to the Planck energy $E_P$. Nonetheless, it exhibits an interesting feature that was not apparent in Eq.~\eqref{VI_rate}: the classical background solution contributes explicitly to the decoherence rate. Indeed, since $\Gamma \propto l_0^2$, the larger is the average separation between the masses in the absence of the gravitational environment, the stronger is the decoherence. This dependence of the decoherence on the background is a remnant of the non-linear nature of the original problem: such a dependence had been noted also in related analyses of decoherence in non-linear QBM (cf. paragraph (V) of~\cite{hu1993quantum}). Within the present study, however, it can be physically interpreted in terms of the tidal nature of gravitational waves. Indeed, the relative change in the arm length of an interferometer induced by a gravitational wave scales as $\Delta L = hL/2$~\cite{pitkin2011gravitational} and is therefore proportional to the baseline length. Thus, the tidal character of gravitational interactions manifests itself even when the system is prepared in a non-classical state. Because such dependence is a distinctive feature of gravitational waves under a general relativistic framework, it may provide a qualitative signature to distinguish gravitational decoherence from more conventional sources, like blackbody radiation or collisions with surrounding gas molecules~\cite{joos1985emergence,gallis1990environmental,schlosshauer2007decoherence}.

\section*{Acknowledgments}
We acknowledge the PNRR PE Italian National Quantum Science and Technology Institute (PE0000023), {the EU EIC Pathfinder project QuCoM (101046973)} and the University of Trieste (Microgrant LR 2/2011). 

\hfill
\appendix
\section{Regularizations}\label{App_Reg}
To regularize the various divergences encountered due to the long wavelength approximation introduced in Sec.~\ref{sec:FNC}, we introduce an exponential cutoff $\Omega_\text{max}$ in the definition of the delta function
\begin{equation}
    \delta(t) = \frac{1}{2\pi}\int_{-\infty}^{\infty}d\omega e^{i\omega t}e^{-|\omega|/ \Omega_\text{max}}
\end{equation}
This makes the Dirac delta a regular and even function with a well defined value at $t=0$. The parity implies that $d^{(2n+1)}\delta(t)/dt^{(2n+1)}|_{t=0}=0$ for all $n$. Furthermore we have the following equalities
\begin{equation}
\begin{cases}
        \delta(0) = \Omega_\text{max}/\pi\,,\\ 
        \frac{d^2}{dt^2}\delta(t)|_{t=0} = -2\Omega_\text{max}^3/\pi\,,\\
        \frac{d^4}{dt^4}\delta(t)|_{t=0} = 24\Omega_\text{max}^5/\pi\,.
\end{cases}
\end{equation}
Applying the same cutoff to the spatial, three-dimensional delta functions we obtain
\begin{equation}\label{eq:cutoffDiracDelta}
\begin{cases}
    \delta^3(0) = \Omega_\text{max}^3/(\pi^2c^3)\\
    \nabla^2\delta^3(\textbf{x})|_0 = -12\Omega_\text{max}^5/(\pi^2c^5)
\end{cases}\,.
\end{equation}

\section{Hamiltonian transformation}\label{app_pwz}
Recall that the Hamiltonian obtained with the Fermi normal coordinates and starting from the Lagrangian $L'$ reads
\begin{align}\label{app_h'}
     \hat H=& \frac{\hat p'_\xi{}^{2}}{2m} +\frac{1}{2}m\omega^2\hat \xi'{}^2+\frac{\pi G}{c^2}\delta^3(0)\acomm{\hat\xi'{}^2}{ \hat p'_\xi{}^2}\nonumber\\
    &+\sqrt{\frac{\pi G}{c^2}}\hat p'_h(0,t)\acomm{\hat p'_\xi}{\hat\xi'}+2\pi\frac{mG}{c^2}(p'_h{}(0,t))^2\xi'{}^2\nonumber\\
    &+\frac{1}{2}\int d^3x \Big(\hat p'_h(\textbf{x},t)^2+c^2\grad \hat h'(\textbf{x},t)^2\Big)\,.
\end{align}
The Hamiltonian can be written as $\hat H = f'(\hat \xi',\hat p'_\xi,\hat h',\hat p_h')$ where $f'$ describes its functional dependence on the canonical variables. In this Appendix we calculate explicitly the new Hamiltonian $\hat H$ from Eq.~\eqref{FH_FinalH} which, in terms of the new canonical variables $(\hat \xi,\hat p_\xi, \hat h, \hat p_h)$,  is given by
\begin{equation}
    \hat H \coloneqq e^{i\hat F/\hbar} f'(\hat \xi,\hat p_\xi,\hat h,\hat p_h)e^{-i\hat F/\hbar}\,,
\end{equation}
where $\hat F = m\sqrt{\frac{\pi G}{c^2}}\hat p_h\hat \xi^2$. Because $\comm{\hat F}{\hat \xi} = \comm{\hat F}{\hat p_h} = 0$, not all terms in the Hamiltonian will be affected by the transformation. Furthermore, we are interested in a perturbative calculation, thus, for consistency, one should not compute the full transformation. Since $\hat F \sim O(\sqrt{G})$ a second order expansion will suffice
\begin{equation}\label{app_pwz:Transf}
    \hat H = f' + \frac{i}{\hbar}\comm{\hat F }{f'} - \frac{1}{2\hbar^2}\comm{\hat F}{\comm{\hat F}{ f'}}+\mathcal{O}(G^{3/2})
\end{equation}
where the dependence of $f'$ on its variables have been suppressed for convenience. Combining the above observations, the only terms which are actually affected by the commutators in~\eqref{app_pwz:Transf} are $\hat p_\xi{}^2/2m$, the interaction $\propto \acomm{\hat\xi}{\hat p_\xi}$ and the contribution $\propto \grad \hat h^2$.
Firstly one has
\begin{align}\label{App_transF1}
    \frac{i}{\hbar}\comm{\hat F }{\hat p_\xi^2/2m} =  - \sqrt{\frac{\pi G}{c^2}}\hat p_h\acomm{\hat \xi}{\hat p_\xi}\,,
\end{align}
where we see that the extra contribution generated cancels the interaction term in the Hamiltonian~\eqref{app_h'}. The second order contribution from this term reads
\begin{align}\label{transf_kin_2}
	-\frac{1}{2\hbar^2}\comm{\hat F}{\comm{\hat F}{\frac{\hat p^2_\xi}{2m}}}& =-\frac{i}{2\hbar}\frac{\pi mG}{c^2}\hat p^2_h(0,t)\comm{\hat \xi^2}{\acomm{\hat \xi}{\hat p_\xi}}\nonumber\\
	&=+2\pi\frac{mG}{c^2}\hat p^2_h(0,t)\hat\xi^2\,.
\end{align}
The interaction Hamiltonian, instead, will transform as
\begin{align}
\frac{i}{\hbar}\comm{\hat F}{\sqrt{\frac{\pi G}{c^2}}p_h(0,t)\acomm{\hat\xi}{\hat p_{\xi}}}\!&=\!\frac{i\pi mG}{\hbar c^2}\hat p^2_h(0,t)\comm{\hat \xi^2}{\acomm{\hat \xi}{\hat p_\xi}}\nonumber\\
&=-\frac{4\pi mG}{c^2}\hat p^2_h(0,t)\hat\xi^2\,,
\end{align}
which, in addition to~\eqref{transf_kin_2}, cancels the interaction term of $O(G)$ in~\eqref{app_h'}. Next one has to calculate
\begin{align}
    &\frac{i}{\hbar}\comm{\hat F}{\frac{c^2}{2}\int d^3x \grad\hat h(\textbf{x},t)^2}\nonumber\\
    &=\frac{i}{\hbar}\comm{m\sqrt{\frac{\pi G}{c^2}}\hat \xi^2\hat p_h(0)}{\frac{c^2}{2}\int d^3x \grad\hat h(\textbf{x},t)^2}\nonumber\\
    &= \frac{i}{\hbar}m\sqrt{\frac{\pi G}{c^2}}c^2\hat \xi^2\int d^3x\comm{\hat p_h(0,t)}{\frac{1}{2}\grad \hat h(\textbf{x},t)^2}\nonumber\\
    &= \frac{i}{\hbar}m\sqrt{\frac{\pi G}{c^2}}\hat \xi^2c^2\int d^3x \grad \hat h(x)\cdot\grad\comm{\hat p_h(0,t)}{\hat h(\textbf{x},t)}\nonumber\\
    &=m\sqrt{\frac{\pi G}{c^2}}\hat \xi^2c^2\int d^3x\grad \hat h(\textbf{x},t)\cdot\grad \delta^3(x) \nonumber\\
    &= -m\sqrt{\frac{\pi G}{c^2}}\hat\xi^2c^2\grad^2\hat h(\textbf{x},t)|_0
\end{align}
where we have used the equal time commutation relation $\comm{\hat p_h(\textbf{x},t)}{\hat h(\textbf{y},t)} =-i\hbar\delta^3(\textbf{x}-\textbf{y})$ and integration by parts. This contribution is the new interaction Hamiltonian generated by the transformation. Finally, the last contribution to the Hamiltonian is given by
\begin{align}\label{App_transf3}
    -\frac{1}{2\hbar^2}&\comm{\hat F}{\comm{\hat F}{\frac{c^2}{2}\int d^3x \grad\hat h(\textbf{x},t)^2}}=\\
    &= \frac{i}{2\hbar}\pi m^2G\hat \xi^4\comm{p_h(0)}{\int d^3x \grad h(\textbf{x},t)\cdot\grad \delta^3(\textbf{x})}\nonumber\\
    &=  \frac{i}{2\hbar}\pi m^2G\hat \xi^4{\int d^3x\grad \comm{p_h(0,t)} { h(\textbf{x},t)}\cdot\grad \delta^3(\textbf{x})}\nonumber\\
    &= \frac{ m^2\pi G}{2}\hat \xi^4{\int d^3x\grad \delta^3(\textbf{x})\cdot\grad \delta^3(\textbf{x})}\nonumber\\
    &= -\frac{ m^2\pi G}{2} \grad^2\delta^3(\textbf{x})|_0\hat \xi^4
    \end{align}
This is an extra, cutoff-dependent contribution to the system Hamiltonian. Combining summing the contributions~\eqref{App_transF1}-\eqref{App_transf3} one obtains the Hamiltonian of the main text. 

\section{Dirac Quantization of Lagrangian $L$}\label{app:DiracQ}
The quantization of the Lagrangian $L$ involves second order time derivatives, and thus the Legendre transform cannot be directly applied. The formal procedure to obtain the Hamiltonian from $L$ is a combination of Ostrogradski's Hamiltonian construction for higher derivative theory, with Dirac theory of constraints~\cite{dirac2013lectures}. We will show, however, that this procedure does not produce $H_{\text{int}} = -L_{\text{int}}$ as assumed in the works~\cite{cho2025non, torovs2024loss,sharifian2024open}. Moreover, even with this procedure, the canonical variables that are assigned to the system and the environment are also not suitable,  for the reasons detailed in Sec.~\ref{subsec:UT}. Therefore, to obtain a suitable separation of the system and environment degrees of freedom a different route has to be followed, such as that of performing an appropriate global unitary transformation, as detailed in the main text. In what follows, we present the derivation of the Hamiltonian, following the standard procedure for Lagrangian with a second order time derivative.   

For a Lagrangian with second order time derivatives, the Ostrogradski procedure~\cite{whittaker1964treatise} introduces an extra canonical variable defined as 
\begin{equation}\label{dirac:defg}
g(\textbf{x},t) = \dot h(\textbf{x},t)\,,
\end{equation}
such that the Lagrangian $L$ reads
\begin{equation}
L = \frac{m}{2}\dot \xi^2 -\frac{1}{2}m\omega^2\xi^2 +m\sqrt{\frac{\pi G}{c^2}}\dot g(0,t)\xi^2 + L_{\text{free}}\,,
\end{equation}
where
\begin{equation}
    L_{\text{free}} = \frac{1}{2} \int d^3x(g^2(\textbf{x},t)-c^2\grad h(\textbf{x},t)^2\big)\,.
\end{equation}
For the set of variables $(\xi,g,h)$, the corresponding conjugate momenta are defined as
\begin{align}
p_\xi &\coloneqq \frac{\partial L}{\partial \dot \xi}\label{mom_ostr_1}\,,\\
p_h(\textbf{x},t) &\coloneqq \frac{\partial L}{\partial g(\textbf{x},t)}-\frac{d}{dt}\frac{\partial L}{\partial \dot g(\textbf{x},t)}\label{mom_ostr_2}\,,\\
p_g(\textbf{x},t) &\coloneqq \frac{\partial L}{\partial \dot g(\textbf{x},t)} \,\label{mom_ostr_3}\,,
\end{align}
The full set of canonical variables is then $(\xi,p_\xi,h,p_h,g,p_g)$ and the Poisson brackets $\acomm{\cdot}{\cdot}$ defined as usual. Computing the derivatives in Eq.~\eqref{mom_ostr_1}-\eqref{mom_ostr_3} we have
\begin{align}
    p_\xi &= m\dot \xi\,,\\ \label{Constr1}
    p_h(\textbf{x},t) &= g(\textbf{x},t) -2m\sqrt{\frac{\pi G}{c^2}}\delta^3(\textbf{x})\xi \dot \xi\,,\\ 
    p_g(\textbf{x},t) &= m\sqrt{\frac{\pi G}{c^2}} \xi^2\delta^3(\textbf{x})\,.\label{Constr2}
\end{align}
We see that while the canonical momenta $p_\xi$ is an independent degree of freedom, $p_h$ and $p_g$ are not. This is because $p_{h}$ and $p_g$ do not involve time derivatives of $h$ or $g$, and can be expressed in terms of the other canonical variables $(\xi,p_{\xi}, g, h)$. As it should be, this still leaves 4-independent degrees of freedom. For these reasons, Eqs.~\eqref{mom_ostr_2}-\eqref{mom_ostr_3} are viewed as constraints encoded in $\phi_1(\mathbf{x})$ and $\phi_2(\mathbf{x})$, with
\begin{align} 
    \phi_1(\textbf{x}) &= g(\textbf{x},t)-p_h(\textbf{x},t)-2\sqrt{\frac{\pi G}{c^2}}\delta^3(\textbf{x})\xi p_\xi \label{constr_1}\,, \\
     \phi_2(\textbf{x}) &= p_g(\textbf{x},t) - m\sqrt{\frac{\pi G}{c^2}}\xi^2\delta^3(\textbf{x}) \label{constr_2}\,,  
\end{align}
 rather than independent dynamical degrees of freedom.  As per Eqs.~\eqref{Constr1} and~\eqref{Constr2}, one should impose $\phi_1 = \phi_2 = 0$. However, Eqs.~\eqref{constr_1} and ~\eqref{constr_2} give the following Poisson brackets 
\begin{equation}\label{second_class}
    \acomm{\phi_1(\textbf{x})}{\phi_2(\textbf{y})} = \delta^3(\textbf{x}-\textbf{y})-\frac{4m\pi G}{c^2}\xi^2\delta^3(\textbf{x})\delta^3(\textbf{y})\neq 0\,,
\end{equation}
which is inconsistent with $\phi_1(\mathbf{x}) = \phi_2(\mathbf{x}) = 0$.
This motivates the introduction of the so-called Dirac brackets $\{\cdot,\cdot\}_D$ which are constructed so that $\{\phi_1,\phi_2\}_D =0$, and is thus consistent with the constraints $\phi_1=\phi_2=0$.

In terms of the standard Poisson brackets, the Dirac brackets for two observables $O_1$ and $O_2$ are defined to be
\begin{align}\label{Dirac_brackets}
    \acomm{O_1}{O_2}_D&= \acomm{O_1}{O_2}\nonumber\\-&\int d\mathbf{x}d\mathbf{y}\acomm{O_1}{\phi_a(\textbf{x})}[\mathrm{C}^{-1}(\textbf{x},\textbf{y})]^{ab}\acomm{\phi_b(\textbf{y})}{O_2}\,,
\end{align}
where $C$ is the matrix built up by the constraints $\mathrm{C}_{ab}(\textbf{x},\textbf{y}) = \acomm{\phi_a(\textbf{x})}{\phi_b(\textbf{y})}$, which is invertible since all the constraints are second class~\cite{dirac2013lectures,galvao1988quantization} as per Eq.~\eqref{second_class}. Explicitly, we have
\begin{equation}\label{InvConstraints}
    [\mathrm{C}^{-1}]^{12}(\textbf{x},\textbf{y}) = -\delta^3(\textbf{x}-\textbf{y}) -\frac{4m\pi G}{c^2}\xi^2\delta^3(\textbf{x})\delta^3(\textbf{y}) 
\end{equation}
which we calculated to $O(G)$, consistently with the perturbative treatment. 
 
The construction of the Hamiltonian then proceeds, but with respect to the Dirac brackets rather than the Poisson ones.  This switch has important consequences: the brackets between the canonical variables $(\xi,p_\xi,h,p_h,g,p_g)$ change. For example
\begin{equation}
    \{\xi,p_\xi\}_D = 1+ \frac{4\pi mG}{c^2}\xi^2\delta^3(0) + O(G^2).
\end{equation}
Even more drastically, we have that
\begin{equation}\label{eq:SENoCommute}
    \acomm{\xi}{g(\mathbf{x},t)}_D = 2\sqrt{\frac{\pi G}{c^2}}\xi \delta^3(\mathbf{x})\,,
\end{equation}
therefore the canonical variables originally defined do not form two commuting sets which can be identified as a system and an environment.

It is, however, possible to find such commuting subsets. Indeed defining
\begin{align}
    p^D_\xi &= p_\xi-2m\sqrt{\frac{\pi G}{c^2}} g(0,t)\xi\label{p_dirac}\\
    p^D_h(\textbf{x},t) & = g(\textbf{x},t) - 2\sqrt{\frac{\pi G}{c^2}}\delta^3(\textbf{x})\xi p_\xi\label{ph_dirac}
\end{align}
one can check using Eq.~\eqref{InvConstraints} that $(\xi,p_\xi^D)$ and $(h,p_h^D)$ form two commuting sets obeying standard canonical commutation relations, with respect to the Dirac brackets~\eqref{Dirac_brackets}~\cite{galvao1988quantization}. However, from Eqs.~\eqref{p_dirac} and~\eqref{ph_dirac}, keeping in mind that $g=\dot{h}$, a direct comparison with Eqs.~\eqref{ph}-\eqref{pxi} of the main text shows that the variables so defined are the set $(\xi',p_\xi',h',p_h')$. For completeness, we derive the Hamiltonian within Dirac quantization, in terms of the conjugate variables $(\xi,p_\xi^D,h, p_h^D)$ and show that it is the same as in Eq.~\eqref{eq:H'}. Indeed, the Hamiltonian following Dirac prescription reads
\begin{align}
    H^D &= \dot\xi p_\xi  + \int d^3 x \Big(g(\textbf{x},t)p_h(\textbf{x},t) + \dot g(\textbf{x},t)p_g(\textbf{x},t)\Big) -L \nonumber\\
    &= \dot\xi p_\xi  + \int d^3 x \Big(g(\textbf{x},t)p_h(\textbf{x},t) + \dot g(\textbf{x},t)p_g(\textbf{x},t)\Big) \nonumber\\
    &\quad-L' -m\sqrt{\frac{\pi G}{c^2}}\Big(\dot g(0,t)\xi^2 +\frac{2}{m}g(0,t)\xi p_\xi\Big) \nonumber\\
      &\overset{\phi_a=0}{=}\dot \xi \Big(p_\xi-2m\sqrt{\frac{\pi G}{c^2}}g(0,t)\xi\Big)+\nonumber\\
      &\qquad\int d^3xg(\textbf{x},t)\Big(g(\textbf{x},t) - 2\sqrt{\frac{\pi G}{c^2}}\delta^3(\textbf{x})\xi p_\xi\Big) -L'\nonumber\\
      &= \dot \xi p_\xi^D+ \int d^3x g(\textbf{x},t)p_h^D(\textbf{x},t) -L'
\end{align}
which is, definitionally, the Hamiltonian defined from $L'$ given in Eq.~\eqref{eq:H'}. 

Thus, in summary, from the perspective of describing open system dynamics, the Dirac prescription either imposes non-commuting relations for system-environment conjugate variables, or renders the conjugate variables the same as those in the primed representation. As we described in Sec.~\ref{sec:oqs}, both these approaches are not suitable for deriving the system's master equation.

\section{Noise and dissipation kernels}\label{App_ker}
The key object governing the reduced dynamics of the system, within the approximation discussed in Sec.~\ref{sec:FNC}, is given by the two-time correlation function of the environmental variable appearing in the coupling Hamiltonian
\begin{equation}\label{B_EQ:alpha}
 \hbar\alpha(t,t') = \frac{m^2\pi G}{c^2}\Tr[c^2\nabla^2 \hat h(\textbf{x},t)\Big|_0c^2\nabla^2 \hat h(\textbf{x},t')\Big|_0\hat\rho_E]
\end{equation}
The gravitational field in Eq.~\eqref{B_EQ:alpha}, consistently with the perturbative treatment, can be calculated according to its free evolution. Firstly we expand the field in modes
\begin{equation*}
    \hat h(\textbf{x},t) = \int \frac{d^3k}{(2\pi)^{3/2}} \sqrt{\frac{\hbar}{2\omega_k}}\Big(\hat a_k e^{i(\textbf{k}\cdot \textbf{x} -\omega_k t)}+ h.c.\Big)\,,
\end{equation*}
so that
\begin{equation}
    c^2\nabla^2 \hat h(\textbf{x},t)\Big|_0 = -\int \frac{d^3k}{(2\pi)^{3/2}}\sqrt{\frac{\hbar}{2\omega_k}}\omega_k^2\Big(\hat a_k e^{-i\omega_k t}+h.c.\Big)\,,
\end{equation}
where the dispersion relation for a massless field $c^2\textbf{k}^2 = \omega_k^2$ has been used. We assume the initial state of the gravitational field is a thermal state $\hat\rho_E(0) = e^{-\hat{H}_E/k_bT}/Z$ so we obtain
\begin{align}\label{alpha_T}
    &\alpha_T(t,t') = \frac{m^2\pi G}{\hbar c^2}\Tr[c^2\nabla^2 \hat h(\textbf{x},t)\Big|_0c^2\nabla^2 \hat h(\textbf{x},t')\Big|_0\hat\rho_E]\nonumber \\
    =&\frac{m^2 \pi G}{c^2}\times\nonumber\\
    &\int \frac{d^3kd^3q}{(2\pi)^3}\frac{\omega_q^2\omega_k^2}{2\sqrt{\omega_k\omega_q}}\Big\{\Tr[\hat a_k\hat a_q^\dagger\hat\rho_E]e^{-i\omega_kt} e^{i\omega_qt'} +h.c.\Big\}\nonumber\\
    =&\frac{m^2G}{16\pi^2c^2}\times\nonumber\\
    &\int d^3k\omega_k^3\Big\{(1+n_T(\omega_k))e^{-i\omega_k(t-t')} +n_T(\omega_k)e^{i\omega_k(t-t')}\Big\}
\end{align}
where we used the standard relation $\Tr[\hat a_k\hat a_q^\dagger\hat\rho_E] = \delta^3(\textbf{q}-\textbf{k})(n_T(\omega_k)+1)$ where $n_T(\omega_k)$ is the thermal mean occupation of the mode $\textbf{k}$. Finally, expressing the integral in spherical coordinates, noting again that $c^2\textbf{k}^2= \omega_k^2$, we obtain
\begin{align}
    \alpha_T(t,t') =& \frac{m^2G}{4\pi c^5}\int _0^{\infty}d\omega\omega^5(2n_T(\omega) +1)\cos \omega(t-t')\nonumber\\
    &-i\frac{m^2G}{4\pi c^5}\int_0^{\infty}d\omega\omega^5\sin \omega(t-t')\,.
\end{align}
We see that, expectedly, the correlation function is stationary, depending on just the time difference $\tau \coloneqq t-t'$. In terms of its frequency components, it has a characteristic supra-ohmic spectrum $\propto \omega^5$.

\subsection{Noise kernel}
The real part of the correlation function is precisely the noise kernel $\Re[\alpha(\tau)]=\mathcal{N}(\tau)$ introduced in the main text and it has contributions coming from both thermal and vacuum fluctuations. The former are finite and can be expressed as
\begin{align}
    \mathcal{N}_T(\tau) =& \frac{1}{2
    \pi}\frac{m^2G}{c^5}\int_0^\infty d\omega\omega^5n_T(\omega)\cos \omega \tau\nonumber\\
    =&\frac{1}{2\pi}\frac{m^2G}{c^5}\int_0^\infty d\omega \frac{\omega^5\cos \omega\tau}{e^{\hbar\omega/k_bT}-1}\nonumber\\
    =&\frac{1}{2}\frac{m^2G}{c^5\hbar}k_bT\frac{d^4}{d\tau^4}\Big\{\frac{\hbar}{\pi k_bT}\int_0^\infty d\omega \frac{\omega\cos \omega\tau}{e^{\hbar\omega/k_bT}-1}\Big\}\,.
\end{align}
The expression in brackets has been singled out since in the high temperature limit $k_bT/\hbar \gg \omega$ it reduces simply to a Dirac delta:
\begin{equation}\label{app_NoiseKer}
    \mathcal{N}_T(\tau)\overset{k_b T \gg \hbar\omega}{=} \frac{1}{2}\frac{m^2 G}{c^5\hbar}k_bT\frac{d^4}{d\tau^4}\delta(\tau)
\end{equation}
The Dirac delta should again be thought of as its regularized counterpart, using the cutoff frequency $\Omega_\text{max}$. The contribution to the noise kernel coming from vacuum fluctuations, instead, reads
\begin{equation}
    \mathcal{N}_0 = \frac{1}{4\pi}\frac{m^2G}{c^5}\int_0^\infty d\omega\omega^5\cos\omega\tau
\end{equation}
is formally divergent; using again the UV cutoff $\Omega_\text{max} \coloneqq 1/\epsilon$, we obtain 
\begin{align}\label{app_NoiseKerVac}
    \mathcal{N}_0(\tau)= \frac{30 m^2G}{\pi c^5} \frac{\epsilon^6 - 15 \epsilon^4 \tau^2 + 
   15 \epsilon^2 \tau^4 - \tau^6}{(\epsilon^2 + \tau^2)^6}\,.
\end{align}

\subsection{Dissipation kernel}
The dissipation kernel $\mathcal{D}$ is instead related to the imaginary component of $\alpha(\tau)$ by
\begin{align}
    \mathcal{D}(\tau) &= -2\theta(\tau)\Im\alpha(\tau)\nonumber\\
    &= \frac{\theta(\tau)}{2\pi}\frac{m^2 G}{c^5}\int_0^\infty d\omega\omega^5\sin \omega\tau
\end{align}
where $\theta$ is Heaviside theta function. Just as the vacuum noise kernel $\mathcal{N}_0$ it requires regularization. If the timescales of interest are large with respect to the inverse of the cutoff $\epsilon$, we can use the functional identity
$\dot\delta(\tau) =-\frac{1}{\pi}\int_0^\infty d\omega \omega\sin \omega\tau$~\cite{parikh2021signatures} to write it simply as
\begin{equation}\label{App_DissKer}
    \mathcal{D}(\tau) = -\frac{\theta(\tau)}{2}\frac{m^2G}{c^5}\frac{d^5}{d\tau^5}\delta(\tau)
\end{equation}
Notice that in the high temperature limit, one can write the fluctuation-dissipation relation in a compact form as 
\begin{equation}
    \mathcal{D}(\tau) = -\theta(\tau)\frac{\hbar}{k_bT}\frac{d}{d\tau}\mathcal{N}_T(\tau)
\end{equation}

\section{Markovian limit of the master equation}\label{APP_markovLim}
We calculate explicitly the coefficients appearing in the master equation~\eqref{PME_MEFull} of the main text, within the limit of timescales $t\gg 1/\Omega_\text{max} = \epsilon$. In the following, to avoid clutter in the equations, we drop the subscript $\hat H_0$ in the Heisenberg evolution of $\hat \xi^2(-\tau)$.
\subsection{Dissipation kernel}
From Eq.~\eqref{ME_nonMarkov} of the main text we have that the contributions due to the dissipation kernel are of the form 
\begin{equation}
    \partial_t\hat\rho = \frac{i}{2\hbar}\int_0^t d\tau \mathcal{D}(\tau) \comm{\hat \xi^2}{\acomm{\hat \xi^2(-\tau)}{\hat \rho}}
\end{equation}
Therefore one needs to evaluate
\begin{equation}
\frac{i}{2\hbar}\int d\tau \mathcal{D}(\tau) \xi^2(-\tau) = -\frac{i}{4\hbar}\frac{m^2G}{c^5}\int_0^t d\tau\frac{d^5}{d\tau^5}\delta(\tau)\hat\xi^2(-\tau)
\end{equation}
The integral can be evaluated by repeated integration by parts. The boundary terms are to be evaluated between $\tau = 0$ and $\tau = t$: since we are interested in times $t\gg \epsilon$ we can assume $\frac{d^n}{d\tau^n}\delta(\tau)|_{\tau = t} =0$; furthermore, at $\tau = 0$, we use the regularized formulas of Appendix~\ref{App_Reg}. We thus have:
\begin{align}
     &-\frac{i}{4\hbar}\frac{m^2G}{c^5}\int_0^t d\tau\frac{d^5}{d\tau^5}\delta(\tau)\hat\xi^2(-\tau) = \nonumber\\
     &=\frac{i}{4\hbar}\frac{m^2 G}{c^5}\Big[\delta^4(\tau)\hat\xi^2(\tau) + \delta^2(\tau) \frac{d^2}{d\tau^2}\hat\xi^2(\tau)+\nonumber\\
     &\qquad\qquad\qquad+\delta(\tau)\frac{d^4}{d\tau^4}\hat\xi^2(\tau) -\frac{1}{2}\frac{d^5}{d\tau^5}\hat\xi^2(\tau)\Big]\Big|_{\tau =0}
\end{align}
Overall therefore, there will be three structurally distinct contributions to the master equation coming from the dissipation kernel:
\begin{align}
    \partial_t\hat\rho =& \frac{i}{\hbar}\frac{m^2 G}{\pi c^5}\Big(6\Omega_\text{max}^5 + \Omega_\text{max}^3\omega^2 + 2\Omega_\text{max}\omega^4\Big)\comm{\hat\xi^4}{\hat \rho}\nonumber\\
    &-\frac{i}{\hbar}\frac{G}{\pi c^5}\Big(\Omega_\text{max}^3 + 2\Omega_\text{max}\omega^2\Big)\comm{\hat\xi^2}{\acomm{\hat p_\xi^2}{\hat\rho}}\nonumber\\
    &-\frac{i}{\hbar}\frac{2mG\omega^4}{c^5}\comm{\hat\xi^2}{\acomm{\acomm{\hat\xi}{\hat p_\xi}}{\hat\rho}}    
\end{align}
The first contribution is a Hamiltonian contribution and, in particular, the term proportional to $\Omega_\text{max}^5$ exactly cancels out the divergence which was brought about by the transformation of Sec.~\ref{sec:oqs} of the main text.

\subsection{Thermal noise kernel}
The contributions from the noise kernel~\eqref{app_NoiseKer}, given the similar distributional form to the dissipation kernel~\eqref{App_DissKer}, can be calculated along the same lines:
\begin{align}
    &\int d\tau \mathcal{N}_T(\tau) \hat\xi^2(-\tau) =\frac{m^2 G k_bT }{2c^5\hbar}\int d\tau\frac{d^4}{d\tau^4}\delta(\tau)\hat\xi^2(-\tau)\nonumber\\
    =&\frac{m^2 Gk_bT}{2c^5\hbar}\times\nonumber\\
    &\Big[-\delta^{(2)}(\tau) \frac{d}{d\tau}\hat\xi^2(\tau) -\delta(\tau)\frac{d^3}{d\tau^3}\hat\xi^2(\tau)+\frac{1}{2}\frac{d^4}{d\tau^4}\hat\xi^2(\tau)\Big]\Big|_{\tau = 0}\nonumber\\
    =&\frac{m^2 G k_bT}{2c^5\hbar}\times\nonumber\\
    &\Big[\Big(\frac{4\Omega_\text{max}\omega^2}{\pi m} + \frac{2\Omega_\text{max}^3}{\pi m}\Big)\acomm{\hat \xi}{\hat p_\xi} -\frac{4\omega^2}{m^2}\hat p_\xi^2+4\omega^4\hat\xi^2\Big]\,.
\end{align}
Thus, the overall contributions are
\begin{align}
    \partial_t\hat\rho_t &= -\frac{1}{\hbar}\frac{m G k_bT}{\pi c^5\hbar}(\Omega_\text{max}^3 + 2\Omega_\text{max}\omega^2)\comm{\hat\xi^2}{\comm{\acomm{\hat\xi}{\hat p_\xi}}{\hat\rho}}\nonumber\\
    +&\frac{1}{\hbar}\frac{2G\omega^2k_bT}{c^5\hbar}\comm{\hat\xi^2}{\comm{\hat p_\xi^2}{\hat\rho}}-\frac{1}{\hbar}\frac{2m^2G\omega^4 k_bT}{c^5\hbar}\comm{\hat\xi^2}{\comm{\hat\xi^2}{\hat\rho}}
\end{align}

\subsection{Vacuum fluctuations kernel}
The contributions coming from the vacuum fluctuations are more involved to compute given the complicated time dependence Eq.~\eqref{app_NoiseKerVac}. However, rewriting the free solution  $\hat \xi^2(-\tau)$ as 
\begin{align}\label{app_xiTrig}
    \hat \xi^2(-\tau) &= \frac{1}{2}\Big(\hat \xi^2 +\frac{\hat p_\xi^2}{m^2\omega^2}\Big) -\frac{1}{2m\omega}\acomm{\hat \xi }{\hat p_\xi }\sin(2\omega \tau)\nonumber\\&+ \frac{1}{2}\Big(\hat \xi^2 -\frac{\hat p_\xi^2}{m^2\omega^2}\Big)\cos(2\omega \tau)\,,
\end{align}
allows us to clearly identify three distinct contributions. Indeed, using the expression~\eqref{app_xiTrig}, we obtain
\begin{align}
    \partial_t \hat\rho|_{\mathcal{N}_0} =& -\frac{f_1(t)}{2\hbar}\comm{\hat\xi^2}{\comm{\hat\xi^2+\frac{\hat p_\xi^2}{m^2\omega^2}}{\hat\rho_t}}\nonumber\\&-\frac{f_2(t)}{2\hbar}\comm{\hat\xi^2}{\comm{\hat\xi^2-\frac{\hat p_\xi^2}{m^2\omega^2}}{\hat\rho_t}}\nonumber\\
    &+\frac{f_3(t)}{2m\hbar\omega}\comm{\hat\xi^2}{\comm{\acomm{\hat\xi}{\hat p_\xi}}{\hat \rho_t}}
\end{align}
The time dependent coefficients $f_i(t)$ can all be evaluated in the limit $t\gg\epsilon = 1/\Omega_\text{max}$.
Firstly we have
$
    f_1(t) = \int_0^t d\tau \mathcal{N}_0(\tau) \overset{t\gg\epsilon}{\longrightarrow}0 
$
Secondly noting that both the cosine and $\mathcal{N}_0(\tau)$, are even functions, we can compute $f_2(t)$  
\begin{align}
    f_2(t)&=\int_0^t ds\mathcal{N}_0\cos2\omega\tau = \frac{1}{2}\int_{-t}^t d\tau\mathcal{N}_0\cos2\omega(\tau) \nonumber\\
    &= \frac{1}{2}\Re \int_{-t}^{t}d\tau \mathcal{N}_0e^{2i\omega\tau}\overset{t\gg \epsilon}{\longrightarrow} \sqrt{\frac{\pi}{2}}\mathfrak{F}[\mathcal{N}_0(2\omega)]\nonumber\\
    &= \frac{1}{4\pi}\frac{m^2G}{c^5}\frac{\pi}{2}(2\omega)^5=\frac{4m^2G\omega^5}{c^5}.
\end{align}
We note that this contribution is cutoff-independent. As is explained in the main text, it describes the quantum noise associated with the spontaneous emission of gravitons. Finally, there is no simple formula to obtain $f_3(t)$, but in the long time limit we have
\begin{align}
    f_3(t) &= \int_0^t d\tau \mathcal{N}_0(\tau)\sin 2\omega\tau \rightarrow \nonumber\\
    &\overset{t\gg\epsilon}{\longrightarrow}-\frac{m^2G}{\pi c^5} \Big[3\omega\Omega_\text{max}^4 + 2\omega^3\Omega_\text{max}^2+\nonumber\\&\qquad\quad -8\omega^5 \ln(2\omega/\Omega_\text{max}) - 8\gamma_{EM}\omega^5\Big]
\end{align}
\subsection{Full master equation}
Putting together all the contributions just calculated we obtain the full Markovian master equation
\begin{widetext}
\begin{align}\label{App_MEFull}
    \partial_t\hat \rho(t) = -&\frac{i}{\hbar}\comm{\frac{\hat p_\xi^2}{2m}+\frac{1}{2}m\omega^2\hat\xi^2 -\frac{m^2 G \Omega_\text{max}\omega^2}{\pi c^5}(\Omega_\text{max}^2+ 2\omega^2)\hat \xi^4+\frac{G \Omega_\text{max}^3}{\pi c^5}\acomm{\hat \xi^2}{p_\xi^2}}{\hat\rho(t)}\nonumber\\
    -&\frac{i}{\hbar}\frac{G \Omega_\text{max}}{\pi c^5}\Big(\Omega_\text{max}^2+2\omega^2\Big)\comm{\hat\xi^2}{\acomm{\hat p_\xi^2}{\hat\rho}}-\frac{i}{\hbar}\frac{2mG\omega^4}{ c^5}\comm{\hat\xi^2}{\acomm{\acomm{\hat\xi}{\hat p_\xi}}{\hat\rho(t)}}\nonumber\\
    -&\frac{1}{\hbar}\frac{m G}{\pi c^5}\Big(\frac{k_bT}{\hbar}\Omega_\text{max}^3+2\frac{k_bT}{\hbar}\Omega_\text{max}\omega^2 +\frac{3}{2}\Omega_\text{max}^4+\Omega_\text{max}^2\omega^2-4\omega^4\log\Big(\frac{2\omega}{\Omega_\text{max}}\Big) -4\gamma_{EM}\omega^4\Big)\comm{\hat\xi^2}{\comm{\acomm{\hat\xi}{\hat p_\xi}}{\hat\rho(t)}}\nonumber\\
    -&\frac{2m^2G\omega^5}{     \hbar c^5}\Big(1+\frac{k_bT}{\hbar\omega}\Big)\comm{\hat \xi^2}{\comm{\hat \xi^2}{\hat \rho(t)}} + \frac{2G\omega^3}{ \hbar c^5}\Big(1+\frac{k_bT}{\hbar\omega}\Big)\comm{\hat \xi^2}{\comm{\hat p_\xi^2}{\hat \rho(t)}}
\end{align}
\end{widetext}
By defining 
\begin{align}
    f(\Omega_\text{max},\omega, T) &= \frac{k_bT}{\hbar}\Omega_\text{max}^3+2\frac{k_bT}{\hbar}\Omega_\text{max}\omega^2 +\frac{3}{2}\Omega_\text{max}^4\nonumber\\
    &+\Omega_\text{max}^2\omega^2-4\omega^4\log\Big(\frac{2\omega}{\Omega_\text{max}}\Big) -4\gamma_{EM}\omega^4 \,,
\end{align}
and
\begin{equation}
g :=\Omega_{\mathrm{max}}(\Omega_\text{max}^2+ 2\omega^2)  
\end{equation}
the expression is the master equation~\eqref{PME_MEFull} of the main text.

\section{Semiclassical dynamics}\label{app:GravBremsstrahlung}
An exact treatment of the dynamics~\eqref{PME_MEFull} is technically challenging. Indeed, the master equation is non-Gaussian and, further, the equations of motion for the moments of $\hat \xi$ and $\hat p_\xi$ form an infinite hierarchy: lower-order moments are coupled to higher-order ones, and the resulting tower of equations does not close. This prevents a closed-form solution for generic initial states.

Motivated by this, in this Appendix we examine Eq.~\eqref{PME_MEFull} under semiclassical regimes, where the use of approximations, like Gaussian truncations, are justified in order to establish some connections with known classical physics. Firstly, one expects that when the state of the system is semiclassical, e.g. a coherent state, the dissipative terms in~\eqref{PME_MEFull} should lead to the classical power loss associated with gravitational bremsstrahlung of~\cite{carroll2019spacetime,maggiore2008gravitational}. Secondly, because of the presence of the backaction of the gravitational environment, the master equation should also encode standard thermodynamical behavior, namely relaxation toward a stationary state determined by the bath temperature.

To investigate both the power loss due to brehmsstralung and relaxation, we need to determine the evolution of the system's energy. 
To that end, we consider the sum of it's kinetic and potential terms $\hat{\mathscr{E}} = \hat{\mathscr{E}}_\text{kin}+ m\omega^2\hat\xi^2/2 $ which, according to~\eqref{Ekin}, reads
\begin{equation}
    \hat{\mathscr{E}} = \hat p_\xi^2/2m + \frac{ G \Omega_\text{max}^3}{\pi c^5}\acomm{\hat p_\xi}{\acomm{\hat p_\xi}{\hat \xi^2}} +\frac{1}{2}m\omega^2\hat\xi^2,
\end{equation}
where we have used the regularized formula for the Dirac delta outlined in Appendix.~\ref{App_Reg}.  We compute its evolution according to~\eqref{PME_MEFull} (see Appendix~\ref{App_EOMs} for the details) and obtain $d\langle\hat{\mathscr{E}}\rangle/dt = \mathrm{Tr}\left(\hat {\mathscr{E}}\dot{\hat{\rho}}_t\right)$ to be 
 \begin{align}\label{VII_AvE}
    \frac{d}{dt}\langle\hat{\mathscr{E}}\rangle = &-\frac{4G\omega^4}{c^5}\langle\acomm{\hat\xi}{\hat p_\xi}^2\rangle +\frac{16 \hbar G\omega^3}{c^5}\Big(1+\frac{k_bT}{\hbar \omega}\Big)\langle\hat{\mathscr{E}}\rangle\nonumber\\
    &+\frac{2mG\Omega_\text{max}\omega^4}{ \pi c^5}\Big(\langle\acomm{\hat\xi^2}{\acomm{\hat\xi}{\hat p_\xi}}\rangle\nonumber\\
    &-\frac{1}{m^2\omega^2}\langle\acomm{\hat p_\xi^2}{\acomm{\hat p_\xi}{\hat\xi}}\rangle\Big)\,.
\end{align}
 We will show that the first contribution represents the irreversible loss of energy due to the emission of gravitational radiation by the system. The second term, instead, is due to the environmental noise, encoding both thermal and vacuum fluctuations of the gravitational field, and drives the system toward relaxation. The last two lines encode the effects of the cutoff $\Omega_\text{max}$. The presence of similar contributions linear in the cutoff has already been encountered in the literature~\cite{kanno2021noise}, however, they do not affect the semiclassical dynamics. Notice, furthermore, that all the terms $\propto \Omega_\text{max}^3$ cancel out exactly. While this does not mean that such terms are irrelevant in general, they do not play a role in the behavior of the system's energy.

Let us firstly focus on just the first contribution of Eq.~\eqref{VII_AvE} which, on classical states, should encode the effects of radiation reaction. In order to make this connection we assume the state to be a highly populated coherent state. Then we can approximate 
$\langle \acomm{\hat \xi}{\hat p_\xi}^2\rangle =  4\langle\hat\xi\rangle^2\langle\hat p_\xi\rangle^2 + \mathcal{O}(\hbar \langle\hat\xi\rangle\langle \hat p_\xi\rangle)$ and  $\langle\acomm{\hat \xi^2}{\acomm{\hat \xi}{\hat p_\xi}}\rangle = 4 \langle\hat \xi\rangle^3\langle \hat p_\xi\rangle +\mathcal{O}(\hbar \langle\hat\xi\rangle\langle \hat p_\xi\rangle)$, with an analogous expression holding for the last term in Eq.~\eqref{VII_AvE}.
Furthermore, since the contributions are already $O(G)$, we can consistently substitute in $\langle\hat \xi\rangle$ and $ \langle\hat p_\xi\rangle$ the background motion, which is just a simple harmonic oscillation. Thus, assuming $\langle\hat \xi\rangle = l_0 \cos \omega t$ with $l_0$ the initial amplitude, we can calculate the average power loss over a period $T=2\pi/\omega$. Since the terms $\propto \Omega_\text{max}$ average to zero over a cycle, we are left with the cutoff-independent terms after averaging over time, which give
\begin{align}\label{semiPower}
    \langle d\langle \hat{\mathscr{E}}\rangle/dt\rangle_T &= -2\frac{G I^2\omega^6}{c^5},
\end{align}
where we have defined the mass quadrupole of the system $I = ml_0^2$. We see that Eq.~\eqref{semiPower} reproduces the standard classical result~\footnote{See e.g. Eq.(3.319) \cite{maggiore2008gravitational}}, apart from the numerical pre-factor. The discrepancy is entirely accounted for by the simplified treatment of the gravitational wave polarizations adopted in the main text. When the polarization modes are properly taken into account, our results match the classical expression exactly, as we show in Appendix~\ref{app:full_pol}.

To examine whether the dynamics reproduces relaxation to equilibrium, we need to further take into account the effects of noise, that is, the second term in~\eqref{VII_AvE}. Let us assume that the system is in a very hot bath $T\gg \hbar \omega$ and that the system is initially prepared in a thermal state as well. Then, we can approximate $\langle \acomm{\hat \xi}{\hat p_\xi}^2\rangle \approx 4\langle\hat\xi^2\rangle\langle\hat p_\xi^2\rangle = 4 \langle\hat{\mathscr{E}}\rangle^2/\omega^2$ while $\langle\acomm{\hat \xi^2}{\acomm{\hat \xi}{\hat p_\xi}}\rangle \approx 0$. Assuming this truncation to hold for the relevant timescales, the evolution takes the simple form
\begin{equation}\label{Energy_ther}
    \frac{d}{dt}\langle\hat{\mathscr{E}}\rangle = -\frac{16G\omega^2}{c^5}\langle\hat{\mathscr{E}} \rangle\Big(\langle\hat{\mathscr{E}} \rangle- k_b T \Big)\,.
\end{equation}
This equation admits a solution describing asymptotic thermalization following 
\begin{equation}\label{DD_therm}
    \langle\hat{\mathscr{E}}(t)\rangle =\frac{ k_bT}{(k_bT/\mathscr{E}_0 -1)e^{-16 \big(\frac{\hbar\omega}{E_p}\big)^2\frac{k_b T}{\hbar}t} +1}\overset{t\rightarrow \infty}{\rightarrow} k_b T 
\end{equation}
where the Planck energy $E_P = \sqrt{\hbar c^5/G}$ has been introduced.  We therefore see that within this regime, the master equation predicts an expected relaxation towards a stationary state of energy $k_bT$. Although Eq.~\eqref{DD_therm} does not describe simple exponential thermalization, we can still identify the relaxation rate $\gamma_R = 16(\hbar\omega/E_P)^2k_bT/\hbar$. Notice that $\gamma_R$ has temperature dependence which is not present in standard QBM and is a byproduct of the non linearity of Eq.~\eqref{Energy_ther}. While this serves to give an estimate of the relaxation rate, its validity is restricted to approximately thermal states. Indeed, because of the non linearity of the opeartors appearing in the master equation~\eqref{PME_MEFull} the timescales governing the evolutions are state dependent.

\section{Equations of motion}\label{App_EOMs}
In this Appendix we derive the equations of motion for the first and second moments of position and momenta, as well as the energy, as predicted by the master equation~\eqref{PME_MEFull}. 
For the average position we have
\begin{equation}\label{app_p}
    \frac{d}{dt}\langle\hat\xi\rangle = \frac{\langle\hat p_\xi\rangle}{m} +\frac{2G\Omega_\text{max}^3}{\pi c^5}\langle\acomm{\hat\xi^2}{\hat p_\xi}\rangle\,.
\end{equation}
The equation of motion for the average momentum is more involved
\begin{align}\label{VII_AvP}
    \frac{d}{dt}\langle\hat p_\xi\rangle &= - m\omega^2\langle\hat\xi\rangle+\frac{4m^2G\Omega_\text{max}\omega^2}{\pi c^5}(\Omega_\text{max}^2+2\omega^2)\langle\hat\xi^3\rangle\nonumber\\
    &-\frac{4G\Omega_\text{max}}{\pi c^5}(\Omega_\text{max}^2+\omega^2)\langle\acomm{\hat p_\xi^2}{\hat\xi}\rangle\nonumber\\
    &-\frac{4mG\omega^4}{c^5}\langle\acomm{\hat\xi}{\acomm{\hat \xi}{\hat p_\xi}}\rangle\nonumber\\
    &-\frac{4\hbar mG}{\pi c^5}f\langle\hat\xi\rangle+\frac{8G\hbar\omega^3}{c^5}\Big(1 + \frac{k_bT}{\hbar\omega}\Big)\langle\hat p_\xi\rangle\,,
\end{align}
where we suppressed the dependence of $f$ on its variables to avoid further clutter. We can now calculate the second moments
\begin{align}\label{App_xi2}
    \frac{d}{dt}\langle\hat \xi^2\rangle=\frac{1}{m}\langle\acomm{\hat \xi}{\hat p_\xi}\rangle +\frac{2G\Omega_\text{max}^3}{\pi c^5}\langle\acomm{\hat\xi^2}{\acomm{\hat\xi}{\hat p_\xi}}\rangle\,.
\end{align}
Then 
\begin{align}\label{App_p2}
    \frac{d}{dt}&\langle\hat p_\xi^2\rangle = -m\omega^2\langle\acomm{\hat \xi}{\hat p_\xi}\rangle\nonumber\\
    &+\frac{2m^2\omega^2G\Omega_\text{max}}{\pi c^5}(\Omega_\text{max}^2+2\omega^2)\langle\acomm{\hat\xi^2}{\acomm{\hat\xi}{\hat p_\xi}}\rangle\nonumber \\
    &-\frac{4G\Omega_\text{max}}{\pi c^5}(\Omega_\text{max}^2+\omega^2)\langle\acomm{\hat p_\xi^2}{\acomm{\hat p_\xi}{\hat \xi}}\rangle\nonumber\\
    &-\frac{8mG\omega^4}{c^5}\langle\acomm{\hat\xi}{\hat p_\xi}^2\rangle\nonumber\\
    &+\frac{16 m^2 \hbar G \omega^5}{ c^5}\Big(1+ \frac{k_bT}{\hbar \omega}\Big)\langle\hat\xi^2\rangle + \frac{16 \hbar G \omega^3}{ c^5}\Big(1+ \frac{k_bT}{\hbar \omega}\Big)\langle\hat p_\xi ^2 \rangle\,,
\end{align}
and finally
\begin{align}\label{VII_Avxp}
    \frac{d}{dt}&\langle\acomm{\hat \xi}{\hat p_\xi}\rangle = \frac{2}{m}\langle\hat p_\xi ^2\rangle -2m\omega^2\langle\hat \xi^2\rangle \nonumber\\
    &+8\frac{m^2 G\Omega_\text{max}\omega^2}{\pi c^5}(\Omega_\text{max}^2+2\omega^2)\langle\hat\xi^4\rangle\nonumber\\
    &-\frac{8mG\omega^4}{c^5}\langle\acomm{\hat \xi^2}{\acomm{\hat \xi}{\hat p_\xi}}\rangle\nonumber\\
    &-\frac{4G\Omega_\text{max}}{\pi c^5}(\Omega_\text{max}^2+2\omega^2)\langle\acomm{\hat\xi^2}{\hat p_\xi^2}\rangle\nonumber\\
    &-\frac{16\hbar mG}{\pi c^5}f\langle\hat \xi^2\rangle+\frac{16\hbar G\omega^3}{ c^5}\Big(1+\frac{k_bT}{\hbar \omega}\Big)\langle\acomm{\hat \xi}{\hat p_\xi}\rangle\,.
\end{align}
As introduced previously the formula for the energy of the system is given by
\begin{equation}
    \hat{\mathscr{E}} =\frac{\hat p_\xi^2}{2m} + \frac{\pi G \delta^3(0)}{c^2}\acomm{\hat p_\xi}{\acomm{\hat p_\xi}{\hat \xi^2}} + \frac{1}{2}m\omega^2\hat\xi^2\,, 
\end{equation}
therefore to assess its evolution, we still need to compute the evolution of $\hat V_G = \frac{\pi G \delta^3(0)}{c^2}\acomm{\hat p_\xi}{\acomm{\hat p_\xi}{\hat \xi^2}}$. Using the canonical commutation relations we can simplify its structure to $\acomm{\hat p_\xi}{\acomm{\hat p_\xi}{\hat\xi^2}} = 2\acomm{\hat p_\xi^2}{\hat\xi^2}+2\hbar^2$ and neglect the constant term. Furthermore for consistency with the perturbative treatment, we have $d\langle\hat V_G\rangle/dt = -\frac{i}{\hbar}\langle\comm{\hat V_G}{\Hat H_S}\rangle + \mathcal{O}(G^2)$. Substituting the free system Hamiltonian we obtain
\begin{align}
    \frac{d}{dt}\langle\hat V_G\rangle =& +\frac{2}{\pi}\frac{G\Omega_\text{max}^3}{mc^5}\langle\acomm{\hat p_\xi^2}{\acomm{\hat p_\xi}{\hat\xi}}\rangle\nonumber\\
    &-\frac{2}{\pi}\frac{G\Omega_\text{max}^3}{c^5}m\omega^2\langle\acomm{\hat\xi^2}{\acomm{\hat\xi}{\hat p_\xi}}\rangle
\end{align}
Thus, using Eqs.~\eqref{App_xi2} and~\eqref{App_p2}  we see that the contributions coming from $d\langle\hat V_G\rangle/dt$ perfectly cancel the terms $\propto\Omega_\text{max}^3$ coming from the rest of the contributions to the energy; we thus obtain
\begin{align}\label{app_E}
    \frac{d}{dt}\langle\hat{\mathscr{E}}\rangle =& -\frac{4G\omega^4}{c^5}\langle\acomm{\hat\xi}{\hat p_\xi}^2\rangle +\frac{16 \hbar G\omega^3}{c^5}\Big(1+\frac{k_bT}{\hbar \omega}\Big)\langle\hat{\mathscr{E}}\rangle\nonumber\\
    &+\frac{2mG\Omega_\text{max}\omega^4}{ \pi c^5}\times\nonumber\\
    &\times\Big(\langle\acomm{\hat\xi^2}{\acomm{\hat\xi}{\hat p_\xi}}\rangle-\frac{1}{m^2\omega^2}\langle\acomm{\hat p_\xi^2}{\acomm{\hat p_\xi}{\hat\xi}}\rangle\Big)\,.
\end{align}
As a final note we report the Wick decompositions of the higher moments present in Eq.~\eqref{app_E}
\begin{align}
    &\langle\acomm{\hat\xi}{\hat p_\xi}^2\rangle = 4\langle\hat\xi^2\rangle\langle\hat  p_\xi^2\rangle +2\langle\acomm{\hat\xi}{\hat p_\xi}\rangle^2  -8\langle\hat p_\xi\rangle^2\langle\hat\xi\rangle^2+\hbar^2\,,\\
    &\langle\acomm{\hat\xi^2}{\acomm{\hat\xi}{\hat p_\xi}}\rangle = 6\langle\hat\xi^2\rangle\langle\acomm{\hat\xi}{\hat p_\xi}\rangle-8\langle \hat p_\xi\rangle\langle\hat\xi\rangle^3\,,\\
    &\langle\acomm{\hat p_\xi^2}{\acomm{\hat p_\xi}{\hat\xi}}\rangle = 6\langle\hat p_\xi^2\rangle\langle\acomm{\hat\xi}{\hat p_\xi}\rangle-8\langle\hat\xi\rangle\langle\hat p_\xi\rangle^3\,,
\end{align}
which are used to analyze the evolution of the energy in semiclassical regimes in Appendix~\ref{app:GravBremsstrahlung}.

\section{Full polarization analysis}\label{app:full_pol}
In the main text we restricted the Lagrangian to capture only the interaction between our two-mass system and gravitational waves traveling in the $z$ direction. This choice was made for clarity of the later analysis; in this Appendix we show that this choice does not affect the analysis up to numerical prefactors in the gravitational kernels of Eqs.~\eqref{noiseKer} and~\eqref{dissKer}.

Let us consider the mode decomposition of the full gravitational field
\begin{equation}
    h_{ij}(\mathbf{x},t)= \sum_{\lambda = +,\times}\int \frac{d^3k}{(2\pi)^{3/2}}h^{\lambda}_\mathbf{k}e_{ij}^{\lambda}(\mathbf{k})e^{i\mathbf{k}\cdot \mathbf{x}}
\end{equation}
where $e_{ij}^\lambda$ are the polarization tensors.  Notice that, since $\hat h_{ij}$ is real we have that $(\hat h^\lambda_{-\mathbf{k}})^* = \hat h^\lambda_{\mathbf{k}}$, while the polarization tensors are chosen to be such that $e_{ij}^{\lambda}(-\mathbf{k}) = e_{ij}^\lambda(\mathbf{k})$, as customary for linear polarizations. Since the two masses are aligned on the $x$ axis, the interaction Lagrangian of Eq.~\eqref{Lint} of the main text reduces to
\begin{align}
    L_{\text{int}} &=\frac{m}{4}\ddot h_{ij}(0,t)\xi^{i}\xi^{j} = \frac{m}{4}\ddot h_{xx}(0,t)(\xi^{x})^2\nonumber\\
    &=\frac{m}{4}\sum_{\lambda = +,\times}\int \frac{d^3k}{(2\pi)^{3/2}}\ddot h^{\lambda}_\mathbf{k}(t)e_{xx}^{\lambda}(\mathbf{k})(\xi^x)^2\,.
\end{align}
By Fourier transforming the mode functions $h_\mathbf{k}^\lambda = \int \frac{d^3x}{(2\pi)^{3/2}}h^{\lambda}(\mathbf{x},t) e^{-i \mathbf{k}\cdot \mathbf{x}}$ we can write the interaction as
\begin{equation}
    L_\text{int} = \frac{m}{4}\sum_{\lambda = +,\times}\int d^3x \ddot h^{\lambda}(\mathbf{x},t)f^{\lambda}(\mathbf{x}) (\xi^x)^2\,,
\end{equation}
where the real functions $f^{\lambda}(\mathbf{x})$ are defined as
\begin{equation}
    f^{\lambda}(\mathbf{x}) = \int \frac{d^3k}{(2\pi)^3}e_{xx}^\lambda(\mathbf{k}) e^{-i\mathbf{k}\cdot \mathbf{x}}\,.
\end{equation}
We see that the fields $h^{\lambda}(\mathbf{x},t)$ do not couple locally to the system, but through an integral over the $f^\lambda(\mathbf{x})$. The full action thus reads
\begin{align}\label{Lprime_full}
    S =& -\frac{c}{2}\sum_{\lambda = +,\times}\int d^4x \partial_\mu h^{\lambda}\partial^{\mu}h^{\lambda}\nonumber\\
    &+\frac{1}{2}\int dt \Big(m\dot \xi^2 -m\omega^2\xi^2\Big)\nonumber\\
    &-2m\sqrt{\frac{\pi G}{c^2}}\sum_{\lambda = +,\times}\int d^3x \dot h^{\lambda}(\mathbf{x},t)f^{\lambda}(\mathbf{x}) \xi\dot \xi
\end{align}
where we have integrated by parts the interaction term and rescaled the gravitational field as in the main text. Notice furthermore that we have dropped the axis label for readability. The full action~\eqref{Lprime_full} is the generalization of the Lagrangian $L'$ used in the main text and we denote the canonical variables in this representation by $(\xi',p'_\xi,h',p_h')$. The conjugate momenta determined by Eq.~\eqref{Lprime_full} read
\begin{align}
    &p'_\xi = m\dot \xi' -2m\sqrt{\frac{\pi G}{c^2}}\sum_{\lambda}\int d^3x \dot {h'}^\lambda(\mathbf{x},t)f^\lambda(\mathbf{x})\xi'\,, \\
   &{p'_h}^{\lambda}(\mathbf{x},t) = \dot{h}'{}^\lambda(\mathbf{x},t)-2m\sqrt{\frac{\pi G}{c^2}}f^\lambda(\mathbf{x})\xi' \dot \xi'\,.
\end{align}
These expressions can be solved for the momenta and the Hamiltonian derived via a Legendre transformation. The result, to the same perturbative order as in Section~\ref{sec:oqs} reads
\begin{align} 
&H = \frac{{p'_\xi}^{\,2}}{2m} +\frac{1}{2}m\omega^2{\xi'}{}^{\,2} \nonumber\\
&\quad+2\sqrt{\frac{\pi G}{c^2}} \sum_{\lambda}\int d^3x\, {p'_h}{}^{\lambda}(\mathbf{x},t) f^{\lambda}(\mathbf{x})\, \xi' p'_\xi \nonumber\\ &\quad +2\pi\frac{mG}{c^2}{\xi'}{}^{\,2} \sum_{\lambda,\mu=+,\times} \int d^3x\,d^3y\, {p'_h}{}^{\lambda}(\mathbf{x}) f^{\lambda}(\mathbf{x}) {p'_h}{}^{\mu}(\mathbf{y}) f^{\mu}(\mathbf{y}) \nonumber\\ &\quad +\frac{2\pi G}{c^2} \sum_{\lambda}\int d^3x\, \bigl(f^{\lambda}(\mathbf{x})\bigr)^2 \xi^2 {p'_\xi}{}^{\,2} \nonumber\\ &\quad +\frac{1}{2}\sum_{\lambda}\int d^3x\, \Bigl( \bigl({p'_h}{}^{\lambda}(\mathbf{x},t)\bigr)^2 +c^2\bigl(\grad {h'}{}^{\lambda}(\mathbf{x},t)\bigr)^2 \Bigr)\,. 
\end{align}
Quantization of the Hamiltonian proceeds as in the main text by imposing canonical commutation relations and Weyl ordering. The Hamiltonian defined above suffers from the same issue discussed in the main text: the canonical variables are not suited for an open quantum systems treatment. To proceed further we need to perform a unitary transformation in order to define a more adequate set of conjugate variables, which we denote by unprimed variables $(\hat \xi,\hat p_\xi,\hat h,\hat p_h)$. The correct analogue of the transformation of Eq.~\eqref{eq:GlobalUnitary} of the main text is given by a unitary $\hat T^\text{full} = e^{\frac{i}{\hbar}\hat F^\text{full}}$ where 
\begin{equation}
    \hat F^\text{full} = m\sqrt{\frac{\pi G }{c^2}}\hat \xi^2\sum_{\lambda=+,\times}\int d^3x f^\lambda(\mathbf{x})\hat p^\lambda_h(\mathbf{x},t)
\end{equation}
Steps closely paralleling to those performed in Appendix~\ref{app_pwz} lead to the following Hamiltonian 
\begin{align}
    \hat H &= \frac{p_\xi^2}{2m}+\frac{1}{2}m\omega^2\hat\xi^2+\frac{1}{2}\sum_{\lambda=+,\times}\int d^3x\Big(p_h^\lambda{}^2 +c^2(\nabla h^\lambda)^2\Big)\nonumber\\
   &-m\sqrt{\frac{\pi G}{c^2}}\sum_{\lambda=+,\times}\int d^3x f^\lambda(\mathbf{x})c^2\nabla^2h{}^\lambda(\mathbf{x},t)\hat \xi^2\nonumber\\
    &+\frac{\pi G}{c^2}\acomm{\hat \xi ^2}{\hat p_\xi {}^2}\sum_{\lambda=+,\times}\int d^3x(f^\lambda(\mathbf{x}))^2\nonumber\\
    &-\frac{\pi Gm^2}{2}\hat\xi^4\sum_{\lambda=+,\times}\int d^3x f^\lambda(\mathbf{x})\grad^2 f^\lambda(\mathbf{x})\,,
\end{align}
where in the first line we collected the unperturbed evolution of the system and gravity, the second line is the coupling Hamiltonian and the final two lines collect the cutoff dependent contributions. Notice that the Hamiltonian of the main text Eq.~\eqref{FH_FinalH} is obtained by restricting to the ``+" polarization only and then setting $f^\lambda(\mathbf{x}) = \delta^3(\mathbf{x})$. We see that the proper treatment of the polarizations results in a non-local coupling, the strength of which is determined by the Fourier transform of the polarization tensors. Now, the field amplitudes $\hat h^\lambda$ behave as two commuting scalar fields, each obeying standard commutation relations with its respective conjugate momentum.  The free evolution of such fields, which is what is relevant for the calculation of the gravitational kernels, is
\begin{equation}\label{mode_exp_new}
    \hat h^\lambda(\textbf{x},t) = \int\! \frac{d^3k}{(2\pi)^{3/2}} \sqrt{\frac{\hbar}{2\omega_k}}\Big(\hat a^\lambda_k e^{i(\textbf{k}\cdot \textbf{x} -\omega_k t)}+ h.c.\Big)\,,
\end{equation}
where $[\hat a^{\lambda}_\mathbf{k},(\hat a^{\lambda'}_\mathbf{q})^\dagger] = \delta^3(\mathbf{k}-\mathbf{q})\delta^{\lambda\lambda'}$. The two point function obtained by tracing over a thermal state of the environment will thus read
\begin{align}
    \hbar \alpha_T^{\text{full}}(t,t') =& \sum_{\lambda=+,\times}\frac{m^2\pi G}{c^2}\int d^3xd^3x'f^{\lambda}(\mathbf{x})f^{\lambda}(\mathbf{x}')\nonumber\\
    &\qquad \times\langle c^2\nabla_\mathbf{x}^2\hat h^\lambda(\mathbf{x},t)c^2\nabla_{\mathbf{x}'}^2\hat h^\lambda(\mathbf{x}',t')\rangle_{\hat\rho_E}
\end{align}
where, comparing with Eq.~\eqref{B_EQ:alpha}, the two point function differs from the original calculation in the presence of the integrals over the $f^{\lambda}(\mathbf{x})$. Using the expansion of Eq.~\eqref{mode_exp_new} and the commutation relations we have
\begin{widetext}
\begin{align}
    &\alpha_T^{\text{full}}(t,t')\! =\frac{m^2 \pi G}{c^2}\!\!\sum_{\lambda=+,\times}\int\!\!d^3xd^3x'f^\lambda(\mathbf{x})f^\lambda(\mathbf{x}')\int\! \frac{d^3kd^3q}{(2\pi)^3}\frac{\omega_q^2\omega_k^2}{2\sqrt{\omega_k\omega_q}}\Big\{\Tr[\hat a^\lambda_\mathbf{k} (\hat a^\lambda_\mathbf{q})^\dagger\hat\rho_E]e^{i(\mathbf{k}\cdot \mathbf{x}-\omega_kt)} e^{-i(\mathbf{q}\cdot \mathbf{x}'-\omega_qt')} +h.c.\Big\}\nonumber\\  
    &=\frac{m^2 \pi G}{c^2}\sum_{\lambda = +,\times}\int d^3xd^3x'f^\lambda(\mathbf{x})f^\lambda(\mathbf{x}')\int \frac{d^3k}{2(2\pi)^3}\omega_k^3\Big\{(1+n_T(\omega_k))e^{i\mathbf{k}\cdot( \mathbf{x}-\mathbf{x}')} e^{-i\omega_k(t-t')}+n_T(\omega_k)e^{-i\mathbf{k}\cdot( \mathbf{x}-\mathbf{x}')} e^{i\omega_k(t-t')} \Big\}\nonumber\\
     &=\frac{m^2 \pi G}{c^2}\int \frac{d^3k}{2(2\pi)^3}\omega_k^3\Big\{(1+n_T(\omega_k)) e^{-i\omega_k(t-t')}  \sum_{\lambda=+,\times}\int d^3xd^3x'f^\lambda(\mathbf{x})f^\lambda(\mathbf{x}')e^{i\mathbf{k}\cdot( \mathbf{x}-\mathbf{x}')}+\nonumber\\
    &\qquad\qquad\qquad\qquad\qquad\qquad+n_T(\omega_k)e^{i\omega_k(t-t')}\sum_{\lambda=+,\times}\int d^3xd^3x'f^\lambda(\mathbf{x})f^\lambda(\mathbf{x}')e^{-i\mathbf{k}\cdot( \mathbf{x}-\mathbf{x}')}\Big\}
\end{align}
\end{widetext}
The integrals over space can be computed
\begin{align}
    \int d^3x f^\lambda(\mathbf{x}) e^{\pm i\mathbf{k}\cdot x} &= \int \frac{d^3q}{(2\pi)^3}e_{xx}^\lambda(\mathbf{q})\int d^3x e^{-i(\mathbf{q}\pm \mathbf{k})\cdot x}\nonumber\\
    &= \int d^3q e_{xx}^\lambda(\mathbf{q})\delta^3(\mathbf{q}\pm\mathbf{k})\nonumber\\
    &= e_{xx}^\lambda(\pm\mathbf{k}) = e_{xx}^\lambda(\mathbf{k})
\end{align}
where we used  the symmetry $e_{xx}^\lambda(\mathbf{k}) = e_{xx}^\lambda(-\mathbf{k})$. Collecting the contributions from the above integrals, we obtain
\begin{align}
    &\alpha_T^\text{full}(t,t')  = \frac{m^2 \pi G}{c^2}\int \frac{d^3k}{2(2\pi)^3}\omega_k^3\Big(\sum_{\lambda} (e_{xx}^\lambda(\mathbf{k}))^2\Big)\nonumber\\
    &\quad\Big\{(1+n_T(\omega_k)) e^{-i\omega_k(t-t')}  +n_T(\omega_k)e^{i\omega_k(t-t')}\Big\}\nonumber\\
    & =\frac{m^2 \pi G}{2{(2\pi)^3} c^5}\int d\omega_k\omega_k^5\times\nonumber\\
    &\Big\{(1+n_T(\omega_k)) e^{-i\omega_k(t-t')} +n_T(\omega_k)e^{i\omega_k(t-t')}.\Big\}\times\nonumber\\
    &\qquad \sum_{\lambda=+,\times}\int  d\Omega_k(e_{xx}^\lambda(\mathbf{k}))^2\,,
\end{align}
where we evaluated the integrals in spherical coordinates, and noted that polarization tensors depend only on the direction of $\mathbf{k}$. Comparing with Eq.~\eqref{alpha_T} we recognise that the integral over $\omega_k$ gives the same time dependence of the original $\alpha(t,t')$, and that the only difference is given by the integration over the solid angle. Therefore we can write
\begin{align}
    \alpha_T^\text{full}(t,t')    &=\frac{\alpha_T(t,t')}{4 \pi}\int d\Omega_k\sum_{\lambda}(e_{xx}^\lambda)(\mathbf{k})^2\nonumber\\
    &=\frac{\alpha_T(t,t')}{4 \pi}\int _0^{2\pi}d\phi\int_{-1}^1\!d\cos\theta (1-\sin^2\theta\cos^2\phi)^2\nonumber\\
    &=\frac{8}{15}\alpha_T(t,t')\,.
\end{align}
We thus see that the original two point function is just rescaled by $8/15$. For example, the formula for brehmsstralung Eq.~\eqref{semiPower}, correcting for this factor gives
\begin{equation}
    \langle d\langle\hat{\mathscr{E}}\rangle/dt\rangle_T = -\frac{16}{15}\frac{G I^2\omega^6}{c^5}\,,
\end{equation}
which matches the correct classical formula~\cite{maggiore2008gravitational,torovs2024loss}.

\bibliography{references_with_doi_url} 
\end{document}